\documentclass[fleqn,usenatbib]{mnras}

\usepackage{amsmath,amssymb, graphicx,xcolor}
\usepackage[export]{adjustbox}
\usepackage{flushend}
\usepackage[utf8]{inputenc}
\usepackage[normalem]{ulem}
\usepackage{multirow}

\usepackage[compact]{titlesec}  
\titlespacing{\section}{0pt}{12pt}{7pt}
\titlespacing{\subsection}{0pt}{9pt}{4pt}

\newcommand{\frb}{FRB~20201124A}

\newcommand{\DM}[1]{#1~pc~cm$^{-3}$}

\newcommand{\RM}[1]{#1~rad~m$^{-2}$}
\newcommand{\Ri}{FRB~20121102A}
\newcommand{\Riii}{FRB~20180916B}
\newcommand{\Rfast}{FRB~20180301A}
\newcommand{\Rbode}{FRB~20200120E}

\usepackage[T1]{fontenc}

\DeclareRobustCommand{\VAN}[3]{#2}
\let\VANthebibliography\thebibliography
\def\thebibliography{\DeclareRobustCommand{\VAN}[3]{##3}\VANthebibliography}

\usepackage{graphicx}	
\usepackage{amsmath}	
\usepackage{amssymb}	

\title[\frb{} polarization]{Polarization properties of \frb{} from detections with the 100-m Effelsberg Radio Telescope}

\author[G. H. Hilmarsson et al.]{
G. H. Hilmarsson$^{1}$\thanks{E-mail: \href{mailto:henning@mpifr-bonn.mpg.de}{henning@mpifr-bonn.mpg.de}},
L. G. Spitler$^{1}$,
R. A. Main$^{1}$,
D. Z. Li$^{2}$
\\
$^{1}$Max-Planck-Institut f{\"u}r Radioastronomie, Auf dem H{\"u}gel 69, D-53121 Bonn, Germany \\
$^{2}$Cahill Center for Astronomy and Astrophysics, MC 249-17 California Institute of Technology, Pasadena CA 91125, USA
}

\date{Accepted XXX. Received YYY; in original form ZZZ}

\pubyear{2021}

\begin{document}
\label{firstpage}
\pagerange{\pageref{firstpage}--\pageref{lastpage}}
\maketitle

\begin{abstract}

The repeating FRB source, \frb{}, was found to be highly active 
in March and April 2021. We observed the source with the 
Effelsberg 100-m radio telescope at 1.36 GHz on 9 April 2021
and detected 20 bursts. 
A downward drift in frequency over time is clearly seen from the majority
of bursts in our sample.
A structure-maximizing dispersion measure (DM) search on the 
multi-component bursts in our sample yields a DM of \DM{$411.6\pm0.6$}.
We find that the rotation measure (RM) of the bursts varies around
their weighted mean value of \RM{-601} with a standard deviation of \RM{11.1}.
This RM magnitude is 10 times larger than the expected Galactic 
contribution along this line of sight (LoS).
We estimate a LoS magnetic field strength of 4--6~$\mu$G, assuming that the entire host galaxy DM contributes to the RM.
Further polarization measurements will help determine \frb{}'s RM stability.
The bursts are highly linearly polarized, with some showing signs
of circular polarization, the first for a repeating FRB. 
Their polarization position angles (PAs)
are flat across the burst envelopes and vary  between bursts. 
We argue that the varying polarization fractions and PAs of \frb{} are similar to  known magnetospheric emission from pulsars, while the observed circular polarization, combined with the RM variability, is hard to explain with Faraday conversion. 
The high linear polarization fractions, flat PAs, and downward drift
from \frb{} bursts are similar to previous repeating sources, while the observed circular polarization is a newly seen behaviour among repeaters.

\end{abstract}

\begin{keywords}
methods: observational -- techniques: polarimetric -- radio continuum: transients
\end{keywords}



\section{INTRODUCTION}

Fast radio bursts (FRBs) are short-duration radio transients
of extragalactic origin. From the roughly 600 FRBs detected so 
far\footnote{\url{https://www.wis-tns.org/}},
repeating bursts have been detected from 24 FRB sources 
\citep{2016Natur.531..202S,2019ApJ...885L..24C,2021arXiv210604352T}.

High levels of activity from a repeating FRB source discovered by CHIME, \frb{}
\footnote{\url{chime-frb.ca/repeaters}},
was announced in March 2021 \citep{2021ATel14497....1C}. 
Active follow-up ensued, where the Australian Square Kilometer Array Pathfinder (ASKAP), the Karl. G. Jansky Very Large Array (VLA), the upgraded Giant Metrewave Radio Telescope (uGMRT), and the European VLBI network (EVN) localized the source to $\sim$arc second precision. 
\citep{2021ATel14515....1D,2021ATel14526....1L,2021ATel14538....1W,2021ATel14603....1M,2021arXiv210611993F}.
Attempts to detect bursts at higher radio frequencies have
yet to be successful \citep{2021ATel14519....1P,2021ATel14537....1S}.

The distance to the host galaxy of \frb{} was determined from 
spectroscopic follow-up observations, which found a redshift of 
$z=0.098 \pm 0.002$ (luminosity distance of 453~Mpc) 
\citep{2021ATel14515....1D,2021ATel14516....1K}.
Persistent radio emission in the host galaxy was detected by 
the uGMRT and the VLA \citep{2021ATel14529....1W,RPP+21}. 
However, \citet{2021arXiv210609710R} find that
the spatial extent and luminosity of
the persistent radio source (PRS) associated with \frb{}
is consistent with star-formation activity,
and place an upper bound on a compact PRS of
$3\times 10^{28}\mathrm{~erg~s}^{-1}\mathrm{~Hz}^{-1}$.

Well studied repeating FRBs, such as \Ri{} and \Riii{},
can provide interesting comparison to the properties of \frb{}.
The rotation measure (RM) of \Ri{} is extremely large, highly variable over time,
consistent with a decade-old supernova remnant (SNR)
\citep{2018Natur.553..182M,2021ApJ...908L..10H}.
\Riii{}'s RM is orders of magnitude lower than \Ri{},
consistent with a several-hundred year old SNR
\citep{2020Natur.577..190M}.
Burst from both \Ri{} and \Riii{} are highly linearly polarized, with
flat polarization angles across the bursts,
and exhibit a downward drift in the burst frequency structure 
\citep{2019ApJ...876L..23H,2019ApJ...885L..24C,2020ApJ...896L..41C}.
Additionally, a periodicity in activity phases has been found for both
\Ri{} and \Riii{} of 160~days \citep{2020MNRAS.495.3551R,2021MNRAS.500..448C}
and 16 days \citep{2020Natur.582..351C}, respectively.

Here we report on our observations of \frb{} with the Effelsberg 100-m radio telescope
in Germany. In \S\ref{sect:obs} we detail our observations and data reduction. 
\S\ref{sect:results} is on our results on the properties of \frb{}. 
In \S\ref{sect:disc} we discuss the implications of our results and compare them
to other repeating FRB sources.
Finally, in \S\ref{sect:summ} we summarize our findings.

\section{OBSERVATIONS AND DATA PROCESSING}\label{sect:obs}

Observations of \frb{} were carried out using the center pixel of the P217mm 7-beam
receiver on the Effelsberg 100-m radio telescope on April 9 2021
starting at 17:46:10 UTC and lasting 4 hours. For the source coordinates (J2000) we used the initial interferometric localization from the ASKAP-CRAFT collaboration: RA = 05h08m03.7s and Dec = +26d03m39.8s \citep{2021ATel14515....1D}. Search mode data were recorded with the Pulsar Fast Fourier Transform Spectrometer backend \citep{2013MNRAS.435.2234B} at a center frequency of 1360 MHz and 300 MHz bandwidth. The search mode data were total intensity filterbank format with 
$512 \times 0.59$ MHz channels and a time resolution of 54.6~$\mu$s. The data were  recorded natively as 32-bit floats and converted to 8-bit unsigned integers prior to searching. Prior to our observations of \frb{} we performed a 3-minute scan of a 1-Hz noise
diode and a 5-minute scan of a test pulsar, B0355+54.

The data were searched for single pulses using
\texttt{HEIMDALL}\footnote{\url{sourceforge.net/projects/heimdall-astro}}, 
\texttt{Presto}\footnote{\url{github.com/scottransom/presto}}\citep{2011ascl.soft07017R}, 
and \texttt{TransientX}\footnote{\url{github.com/ypmen/TransientX}}
from dedispersed time-series between 350 and \DM{450} for boxcar widths
up to 30~ms, down to a signal-to-noise (S/N) threshold of 7.
At this threshold we are sensitive to bursts down to a fluence
of $0.14\times(W_\mathrm{ms})^{1/2}\mathrm{~Jy~ms}$, where $W_\mathrm{ms}$
is the burst width in milliseconds.
For \texttt{Presto} and \texttt{TransientX} a DM step of \DM{1} was chosen,
while \texttt{HEIMDALL} determined its own steps based on pulse broadening
induced by the DM step size.
\texttt{HEIMDALL} missed one of the weaker bursts in our sample, \texttt{Presto} missed a strong burst due to it being in a segment of data flagged as radio frequency interference in the \texttt{rfifind} mask, and \texttt{TransientX} detected all the bursts in our sample.

Voltage data with 1.56~ns time resolution and two polarizations were recorded simultaneously with the search mode data. Around each burst detected in the single pulse search, a 20~s-long segment of data were saved. 
Coherently dedispersed
\texttt{PSRCHIVE}\footnote{\url{psrchive.sourceforge.net}}
\citep{2004PASA...21..302H}
archives containing bursts with full Stokes information
were created from the baseband data using 
\texttt{DSPSR}\footnote{\url{dspsr.sourceforge.net}}
\citep{2011PASA...28....1V}
and were used to perform all our analysis.
The archives were polarization calibrated with a 3-minute scan
of a 1-Hz noise diode performed prior to our observations
using \texttt{PSRCHIVE}'s \texttt{pac}.
We validated our polarization calibration by reconstructing 
the polarization profile of our test pulsar, B0355+54, as
it is in the EPN database of pulsar 
profiles\footnote{\url{epta.eu.org}}, as well as obtaining the 
same RM (\RM{79}) as is listed in the ATNF pulsar 
catalogue\footnote{\url{atnf.csiro.au}} \citep{2005AJ....129.1993M}.

We opted for a structure-maximizing method of determining DMs
instead of S/N maximizing due to the latter leading to overlapping
sub-components and displaying a broader range of apparent DMs
\citep{2019ApJ...876L..23H}.
DMs were determined using 
\texttt{DM\_phase}\footnote{\url{github.com/danielemichilli/DM_phase}}
\citep{2019ascl.soft10004S},
which performs a Fourier transform on the time-series of each frequency channel
over a range of trial DMs and then integrates over the emission frequency
to produce a coherence spectrum.
Fourier frequencies that have similar phase angles will sum to
a greater amplitude. This method works better for bursts with 
sub-components, as the coherent sum will be even greater.

The RMs were determined with RM synthesis 
\citep{1966MNRAS.133...67B,2005A&A...441.1217B}
using \texttt{rmsynth1d} and \texttt{rmclean1d} in the
\texttt{RM-Tools} package\footnote{\url{github.com/CIRADA-Tools/RM-Tools}}\citep{2020ascl.soft05003P} for 1000 trial RMs from -2500 to \RM{2500}.
The RMs are obtained by reconstructing the Faraday dispersion function (FDF)
through a Fourier transform of the Stokes $Q$ and $U$ parameters. 
The width of the FDF peaks rescaled by their maximum S/N 
represents the 1$\sigma$ RM uncertainties.
We compared our RMs to results obtained with \texttt{PSRCHIVE}'s \texttt{rmfit}
and a Bayesian Lomb-Scargle periodogram RM search 
method, \texttt{RMcalc}\footnote{\url{gitlab.mpifr-bonn.mpg.de/nporayko/RMcalc}}
\citep{2019MNRAS.483.4100P}
and saw consistent values across methods.

\section{BURST PROPERTIES AND ANALYSIS}\label{sect:results}

\begin{figure*}
\centering
\includegraphics[width=.24\textwidth]{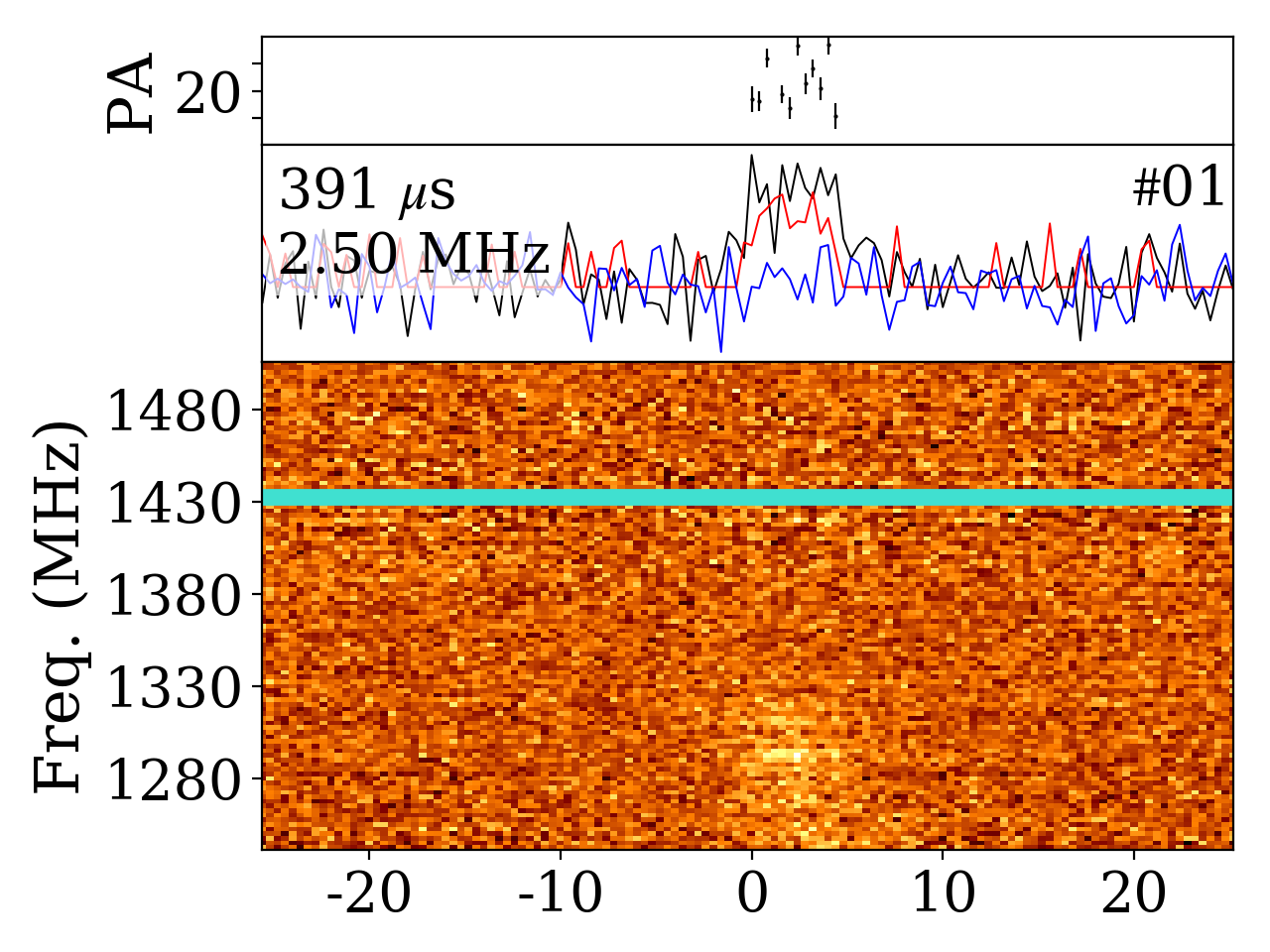}
\includegraphics[width=.24\textwidth]{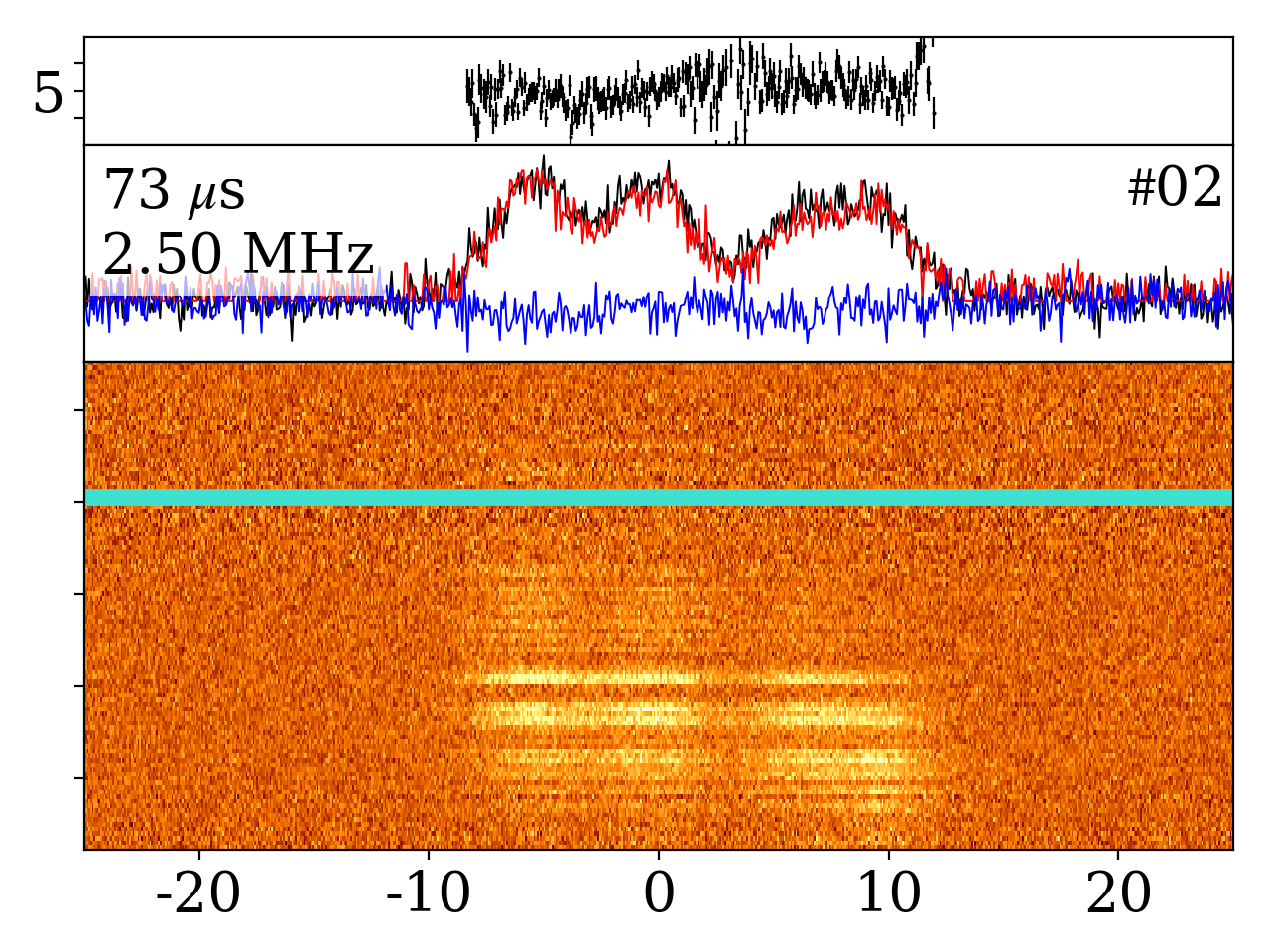}
\includegraphics[width=.24\textwidth]{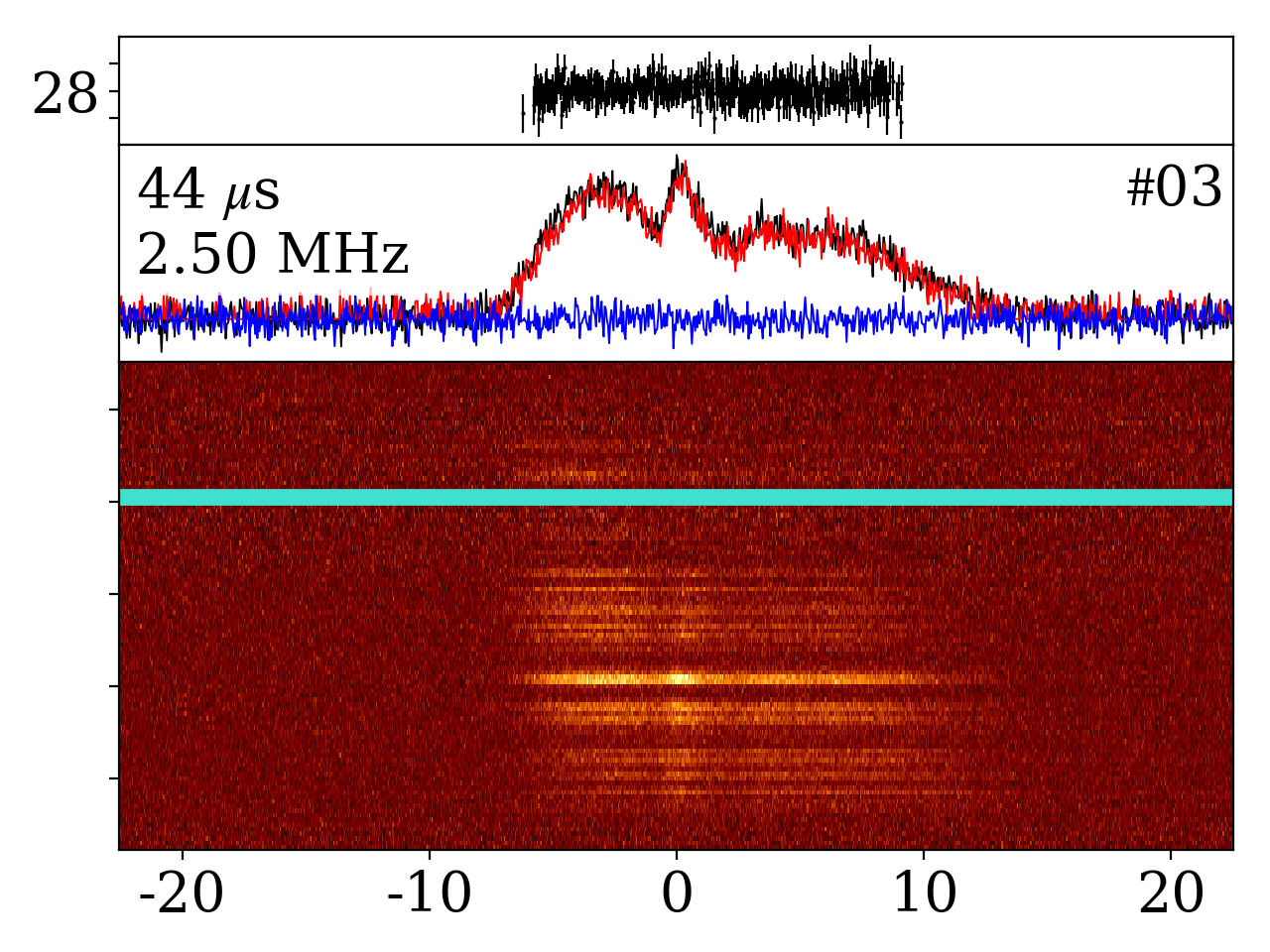}
\includegraphics[width=.24\textwidth]{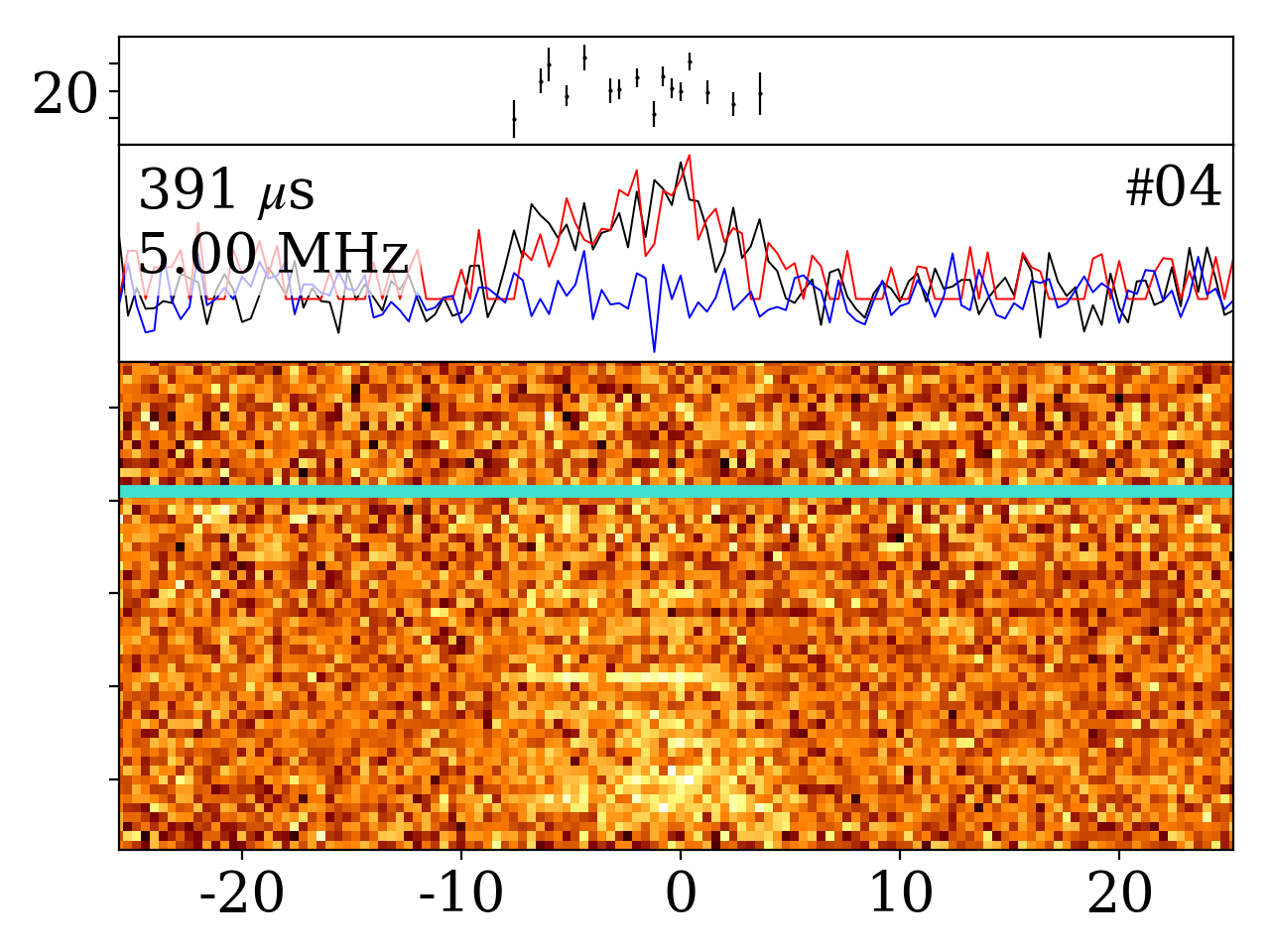}
\includegraphics[width=.24\textwidth]{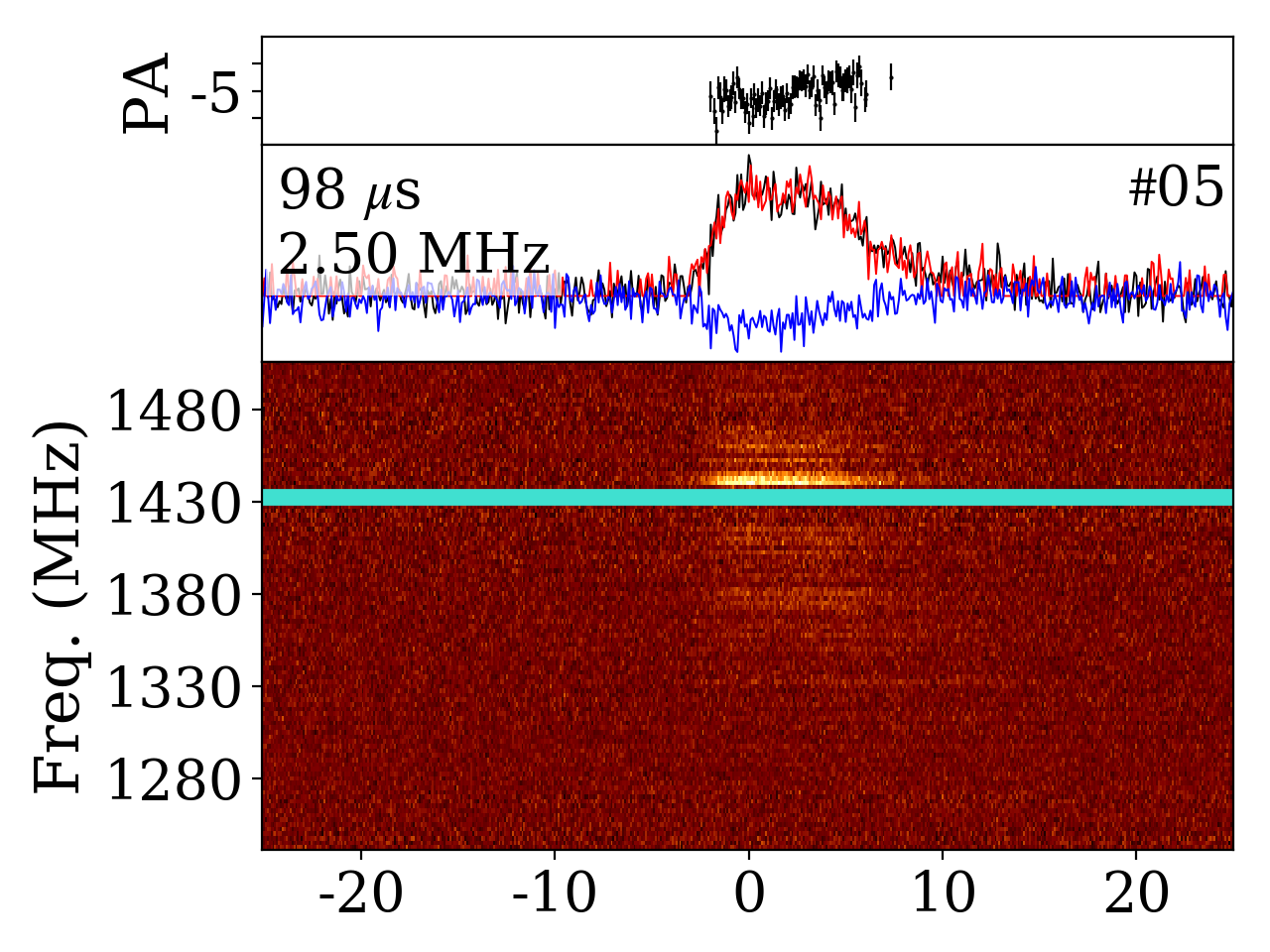}
\includegraphics[width=.24\textwidth]{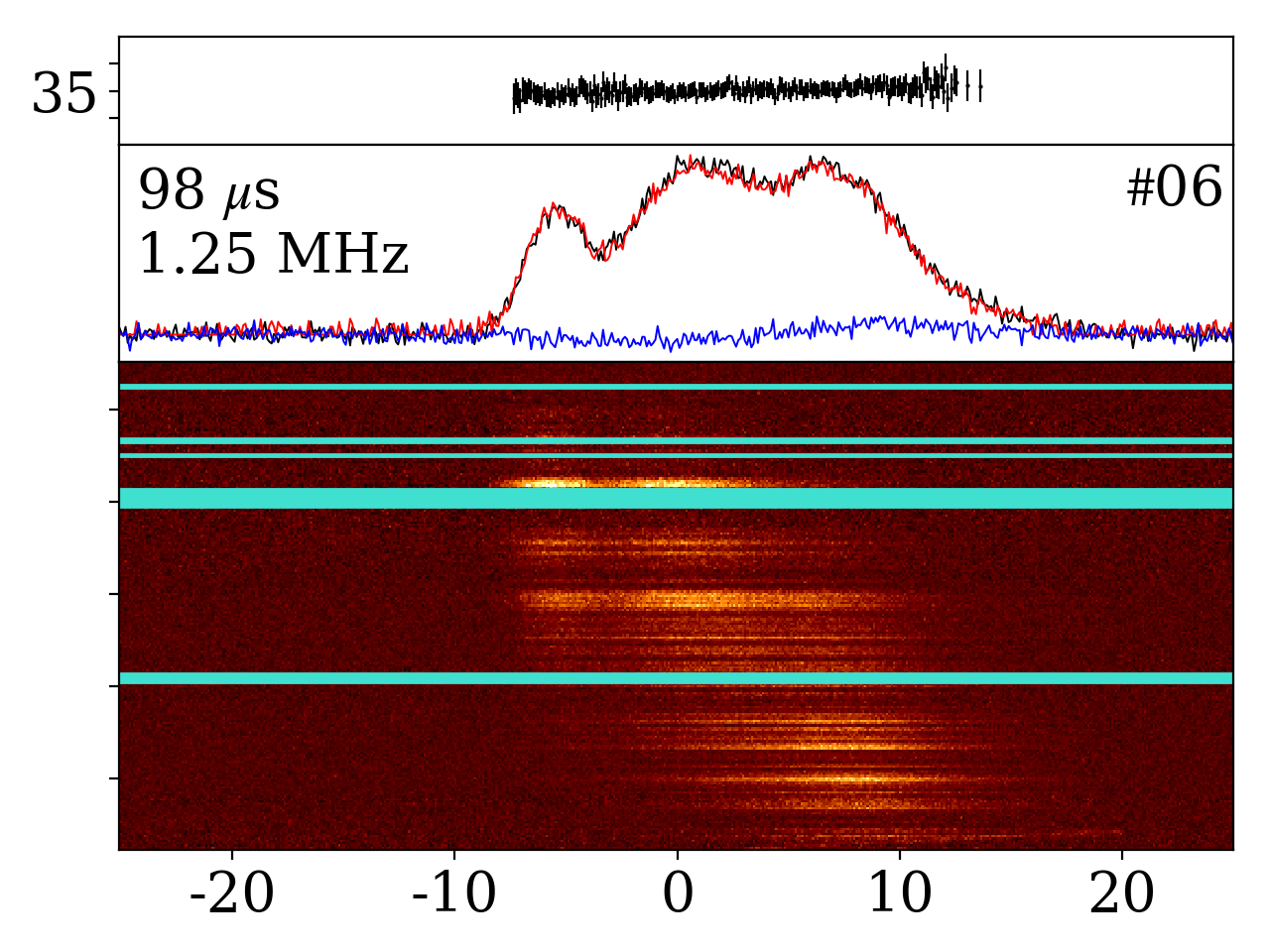}
\includegraphics[width=.24\textwidth]{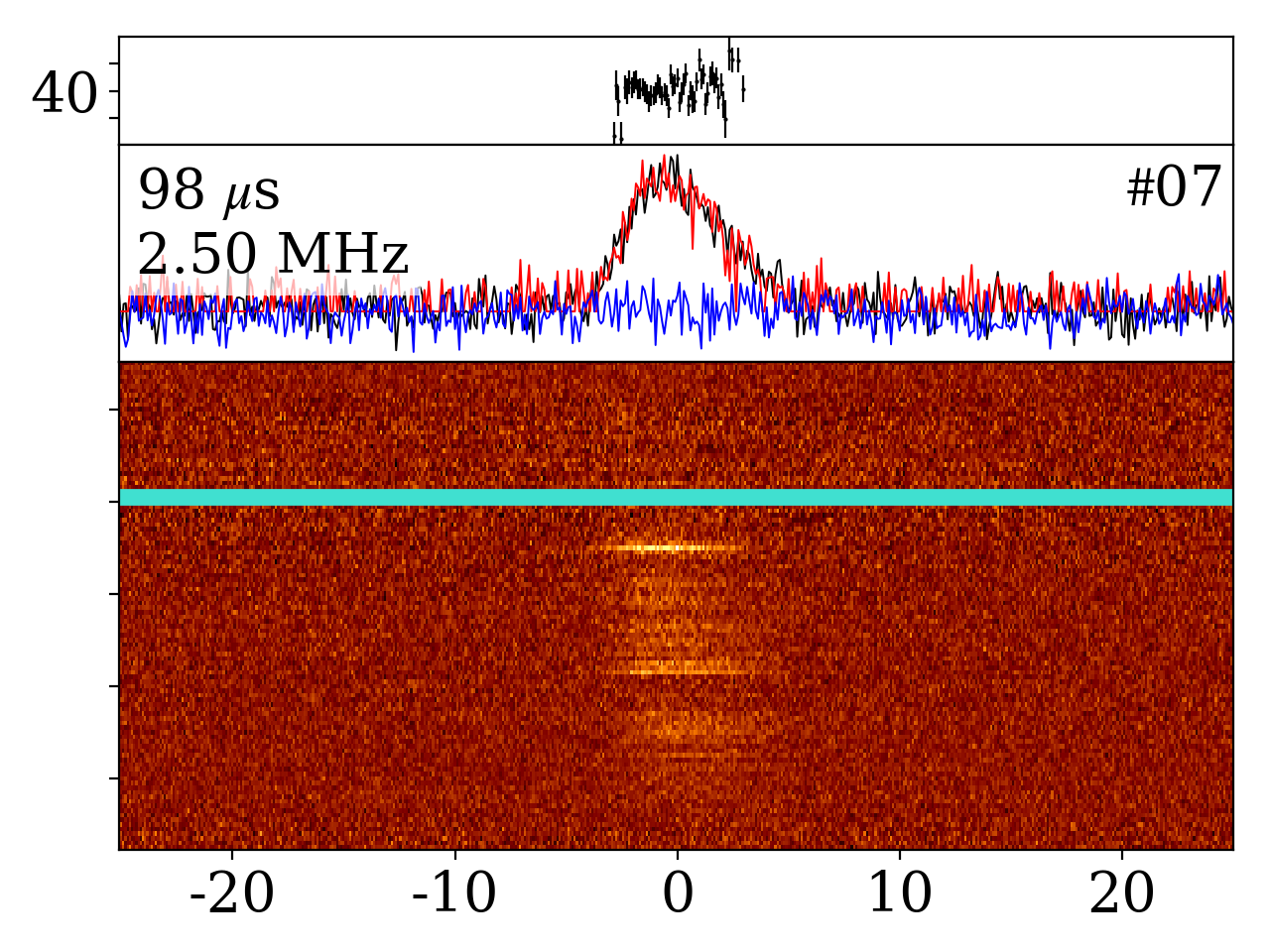}
\includegraphics[width=.24\textwidth]{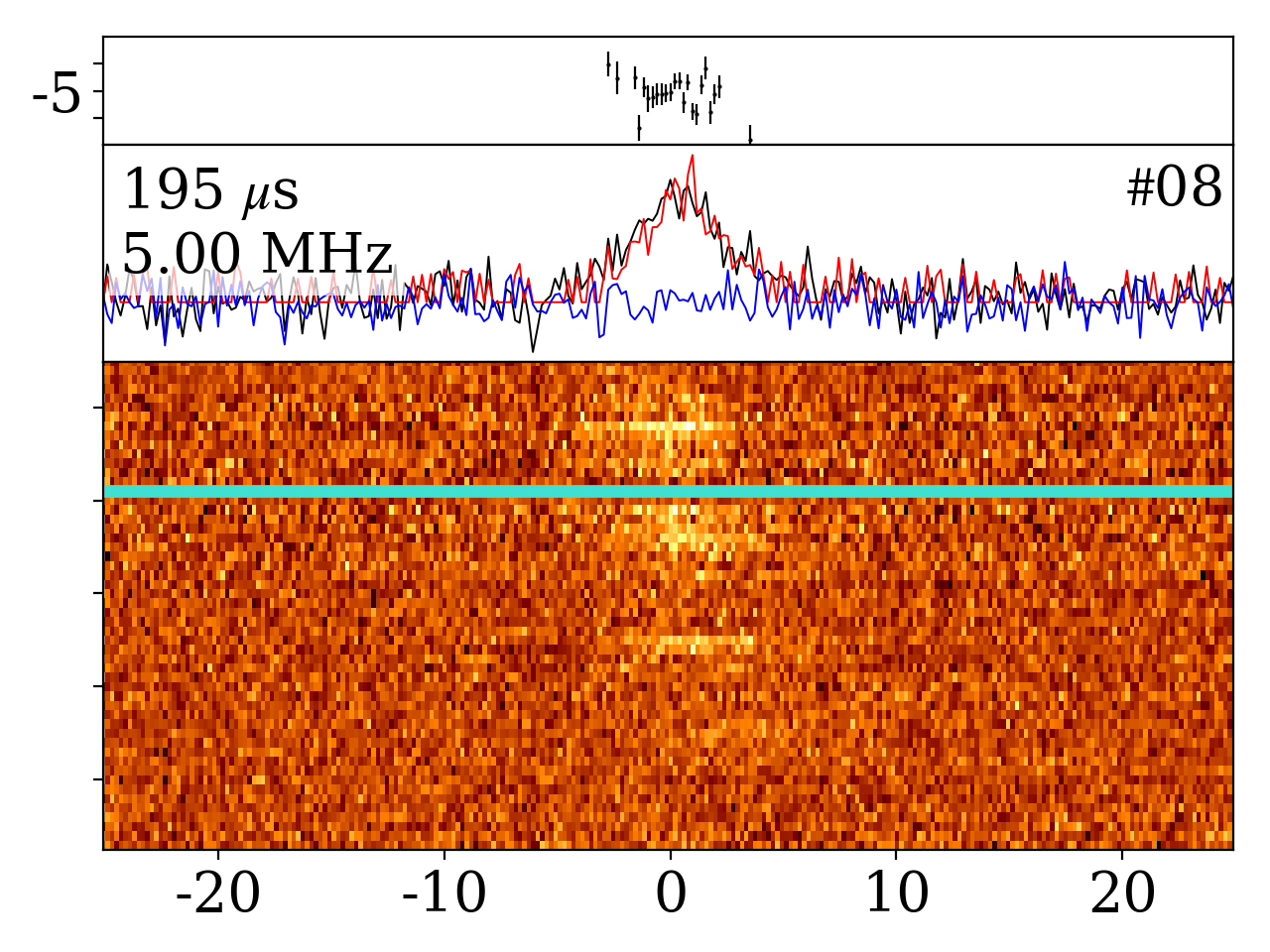}
\includegraphics[width=.24\textwidth]{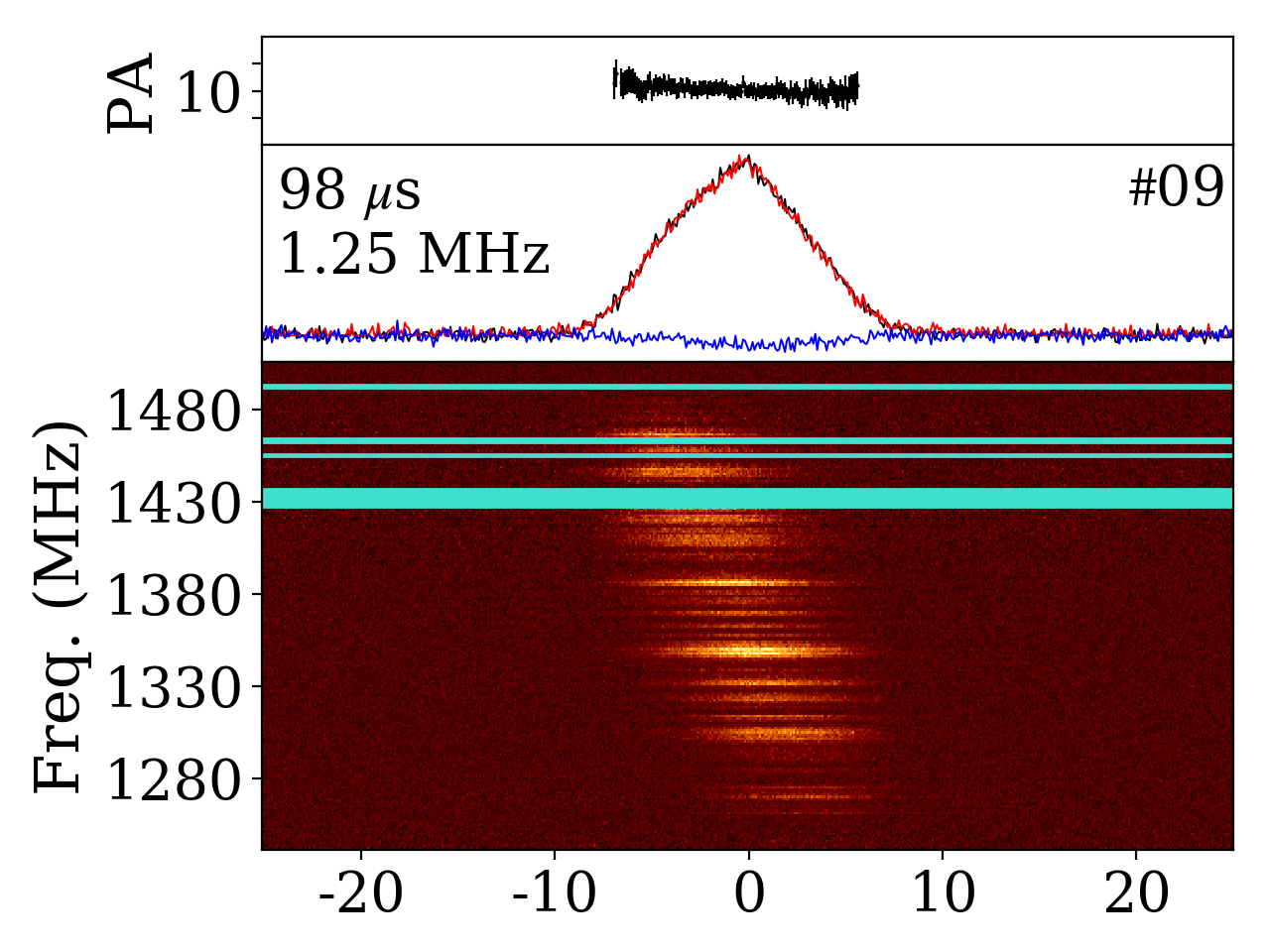}
\includegraphics[width=.24\textwidth]{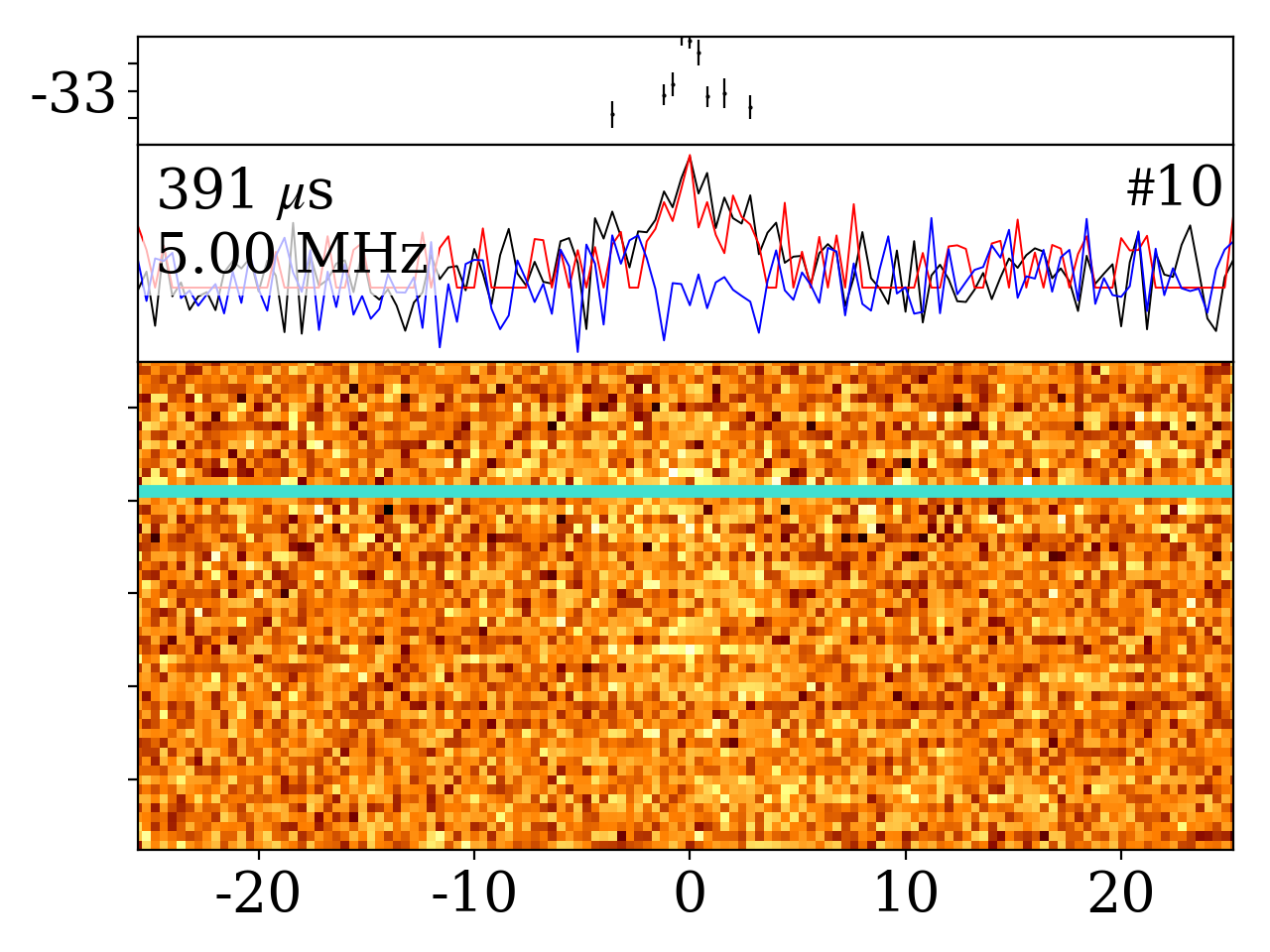}
\includegraphics[width=.24\textwidth]{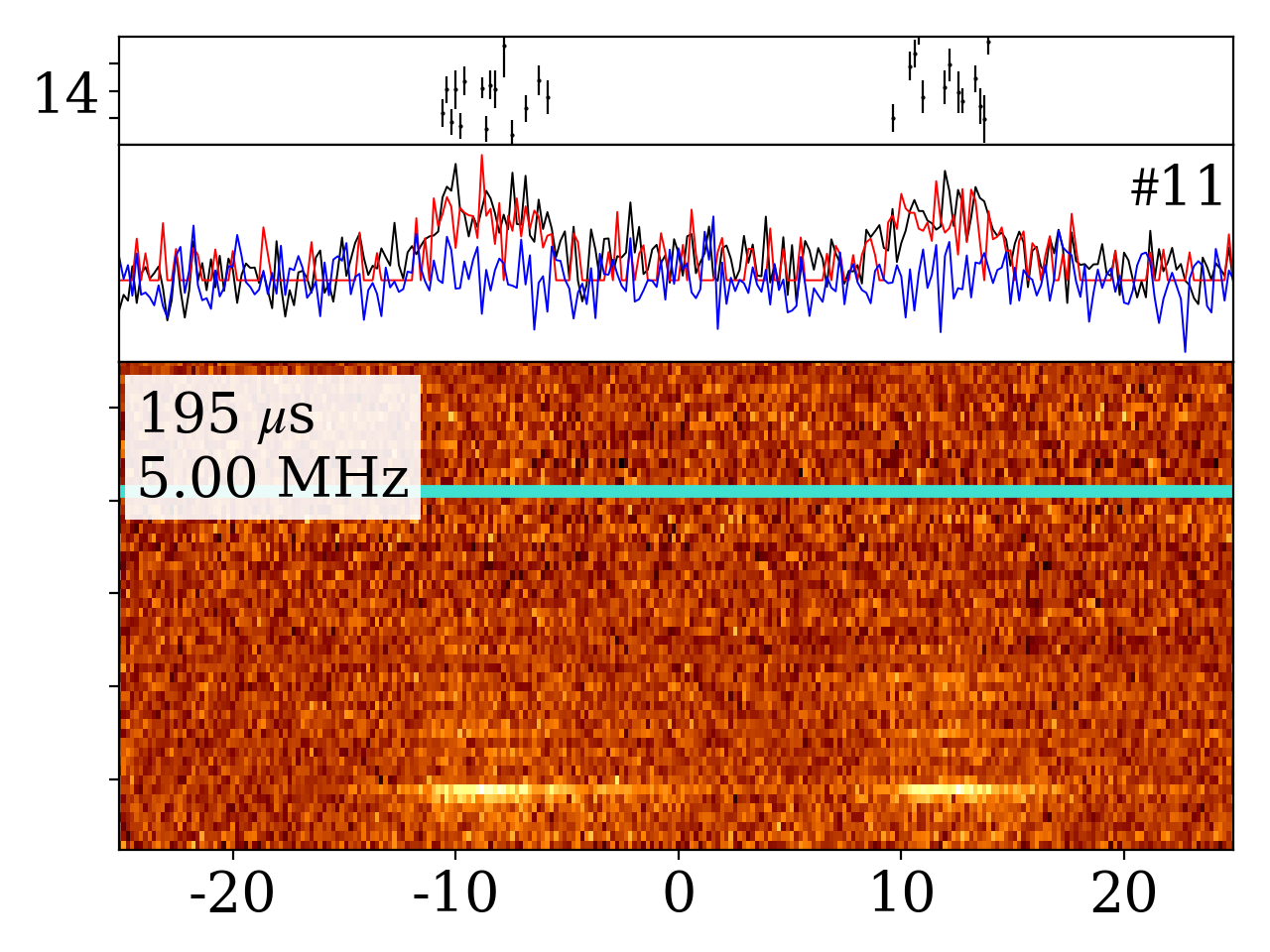}
\includegraphics[width=.24\textwidth]{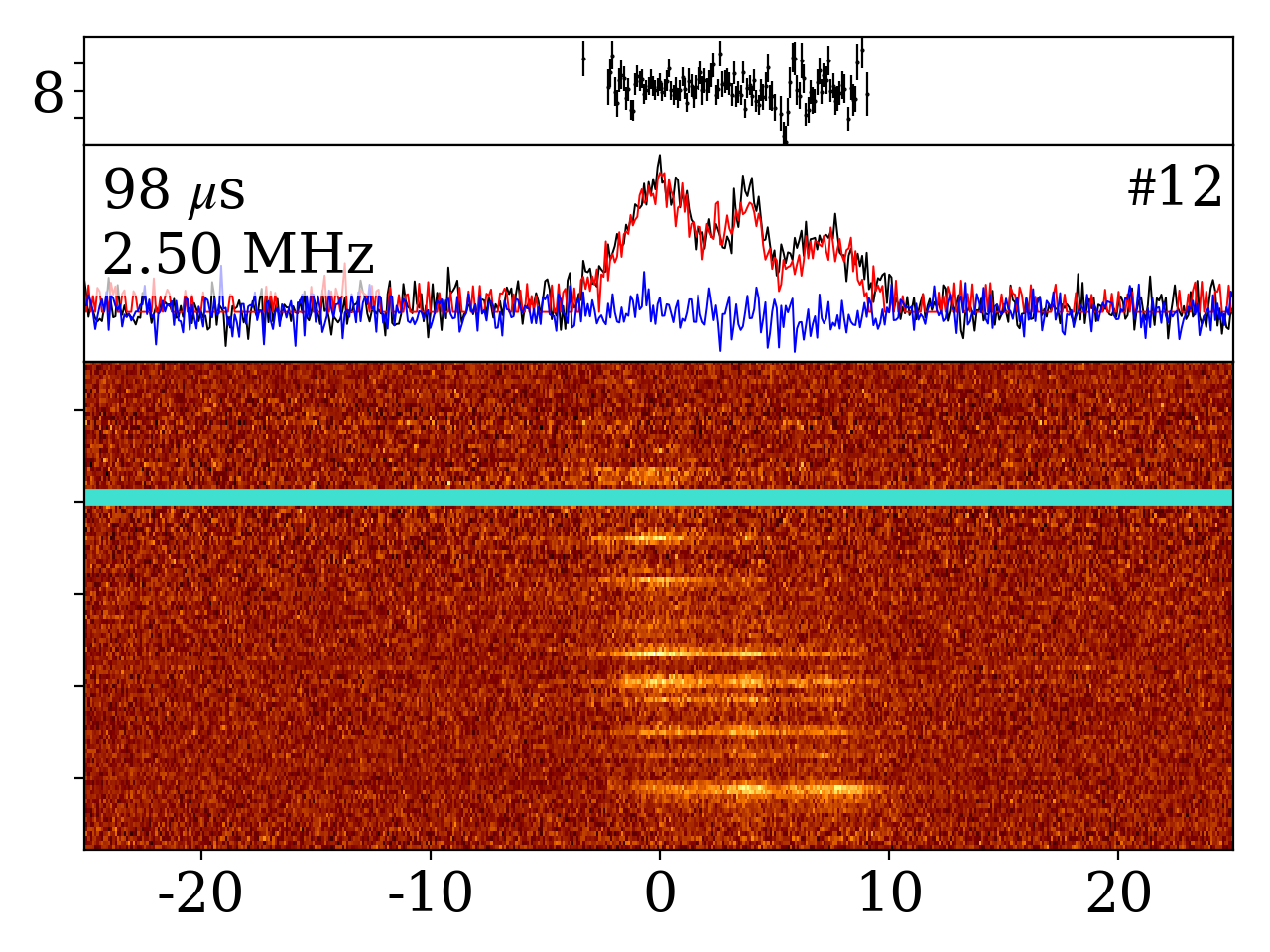}
\includegraphics[width=.24\textwidth]{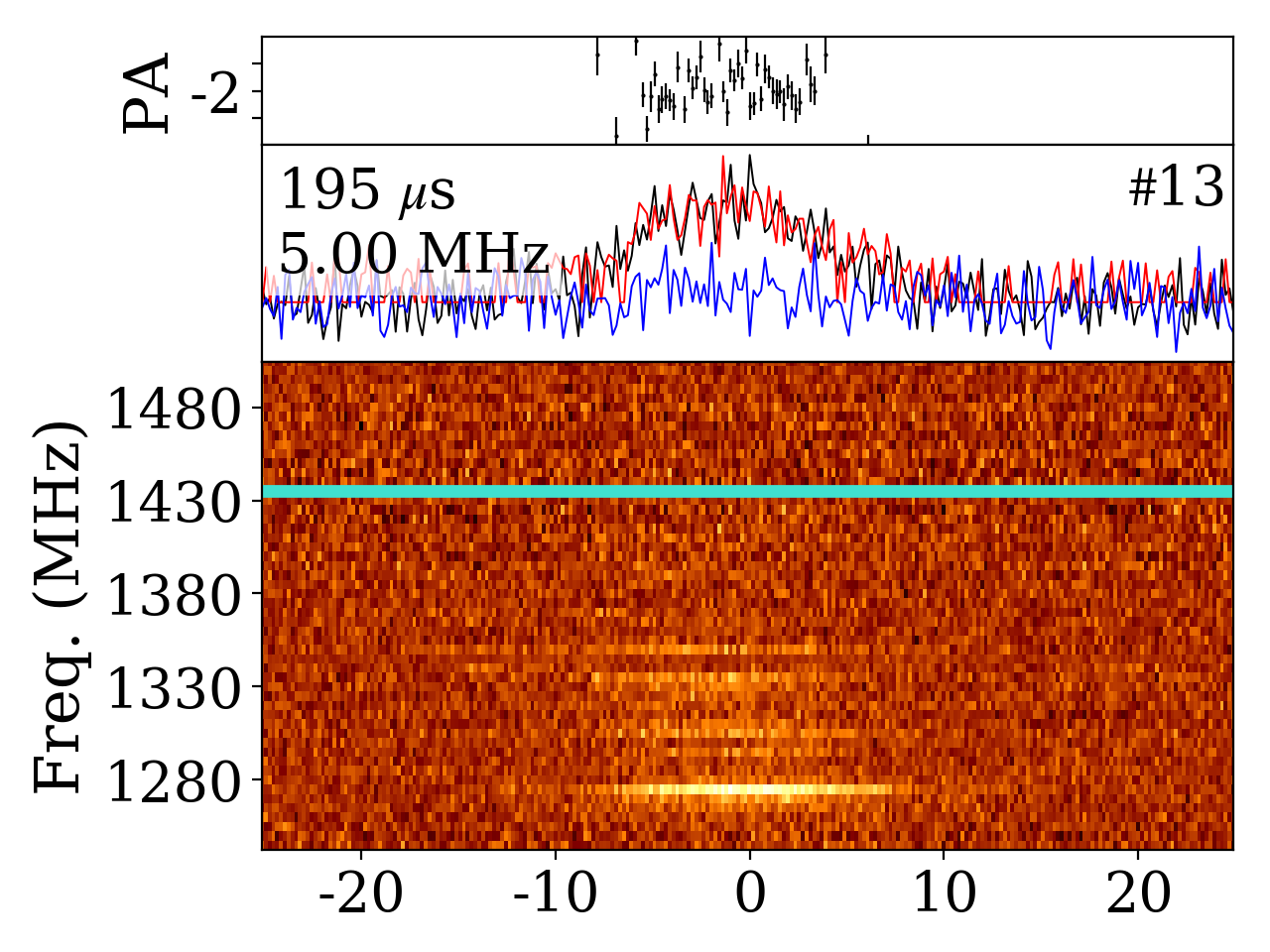}
\includegraphics[width=.24\textwidth]{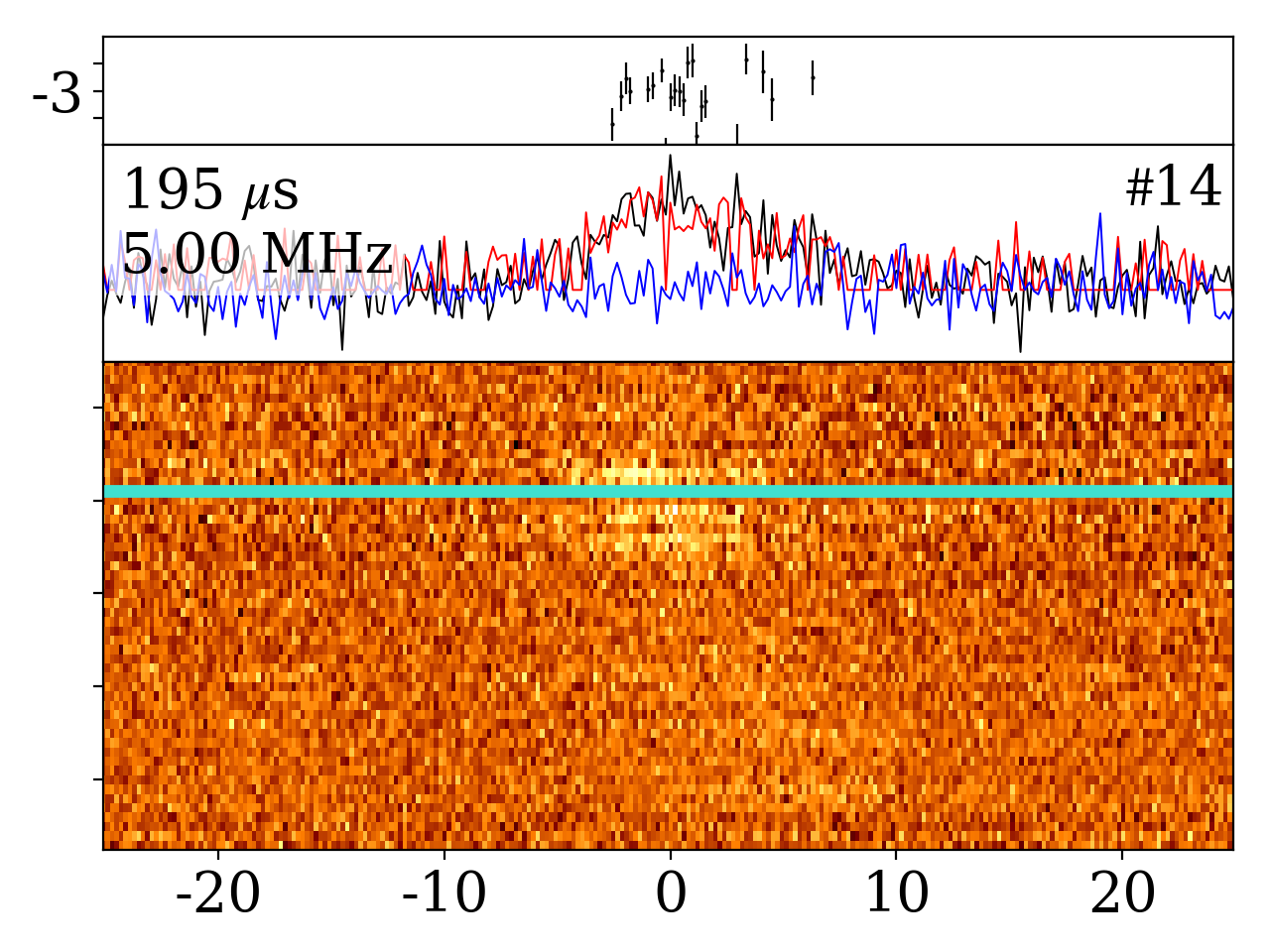}
\includegraphics[width=.24\textwidth]{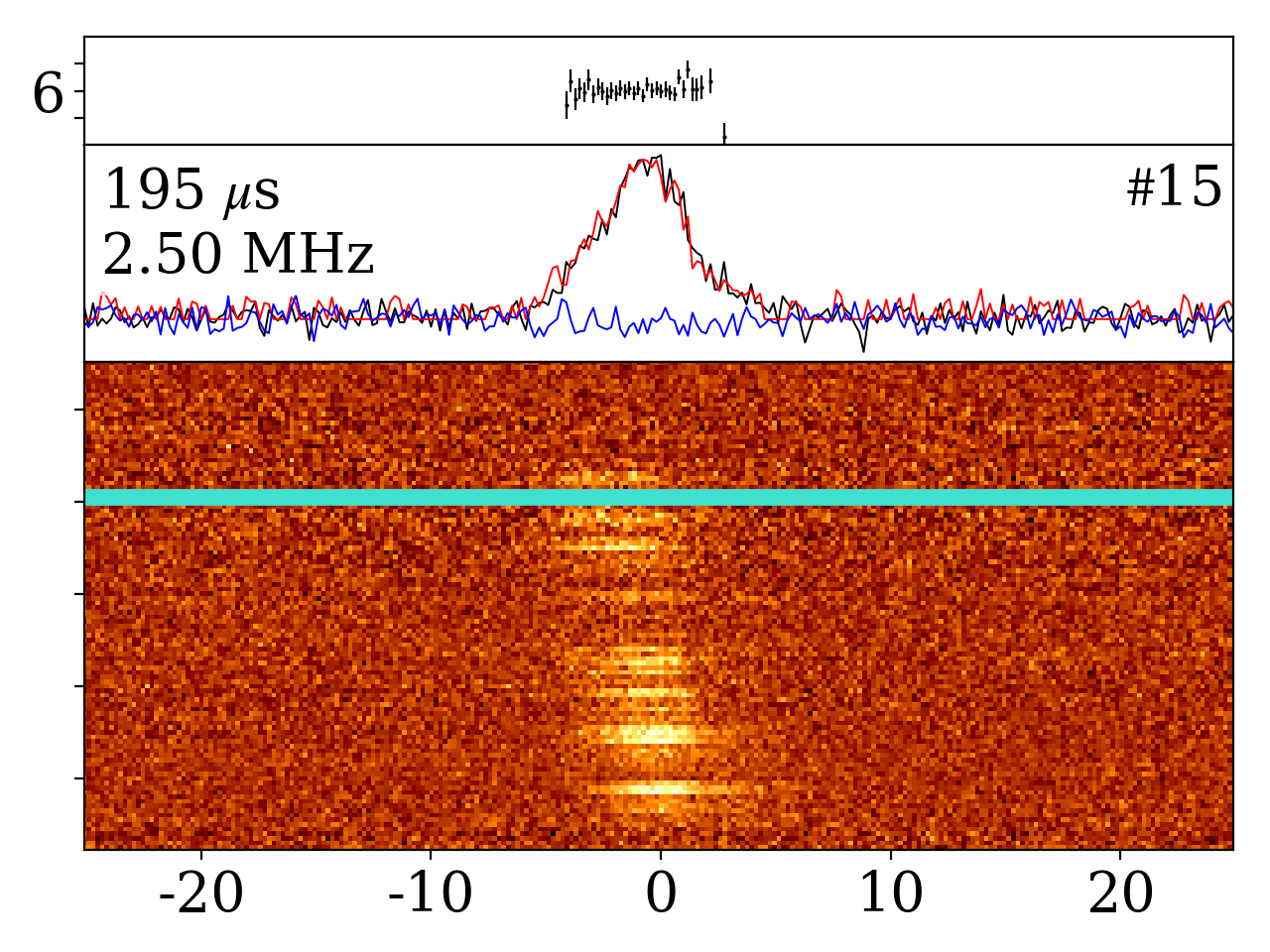}
\includegraphics[width=.24\textwidth]{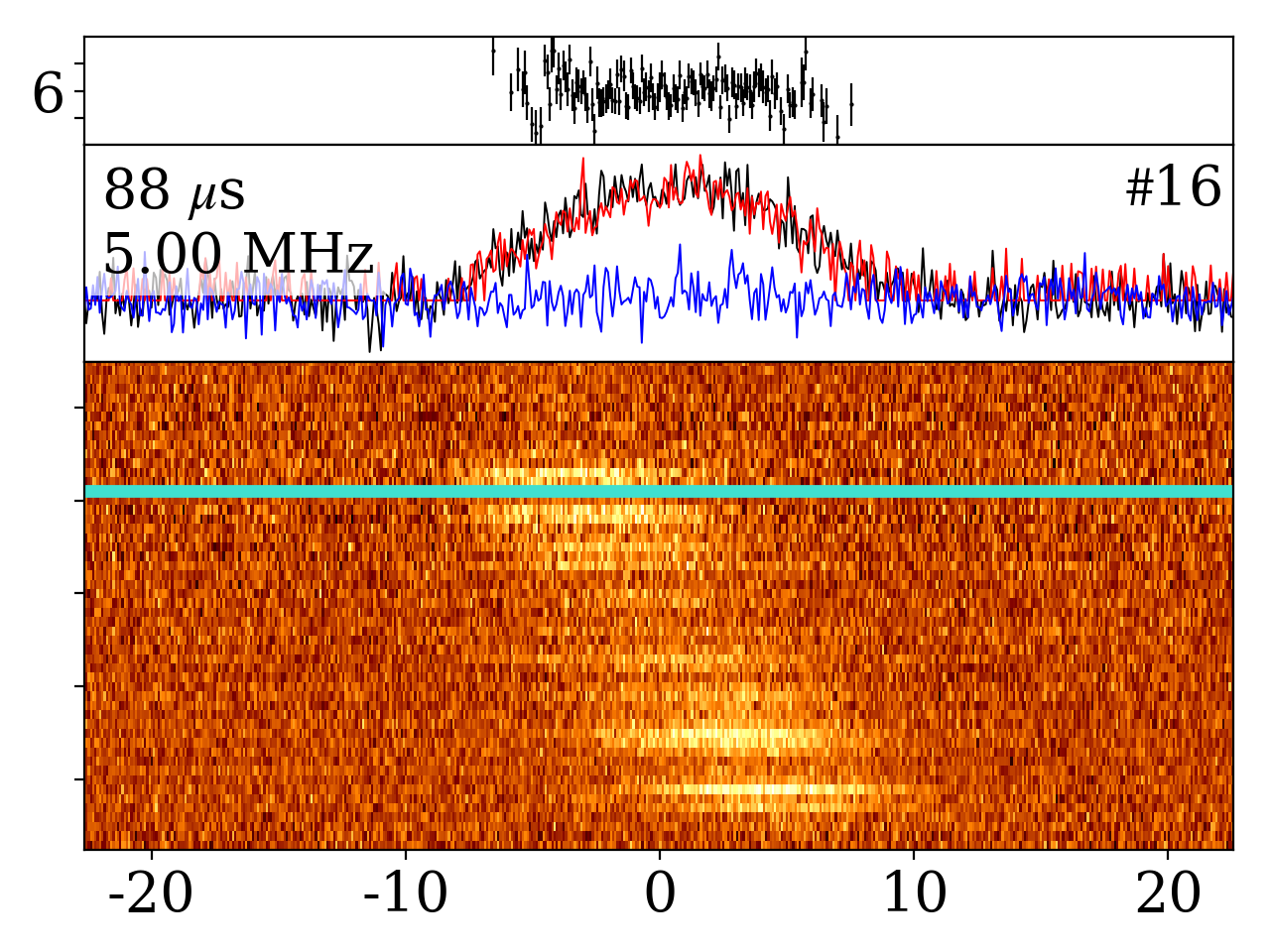}
\includegraphics[width=.24\textwidth]{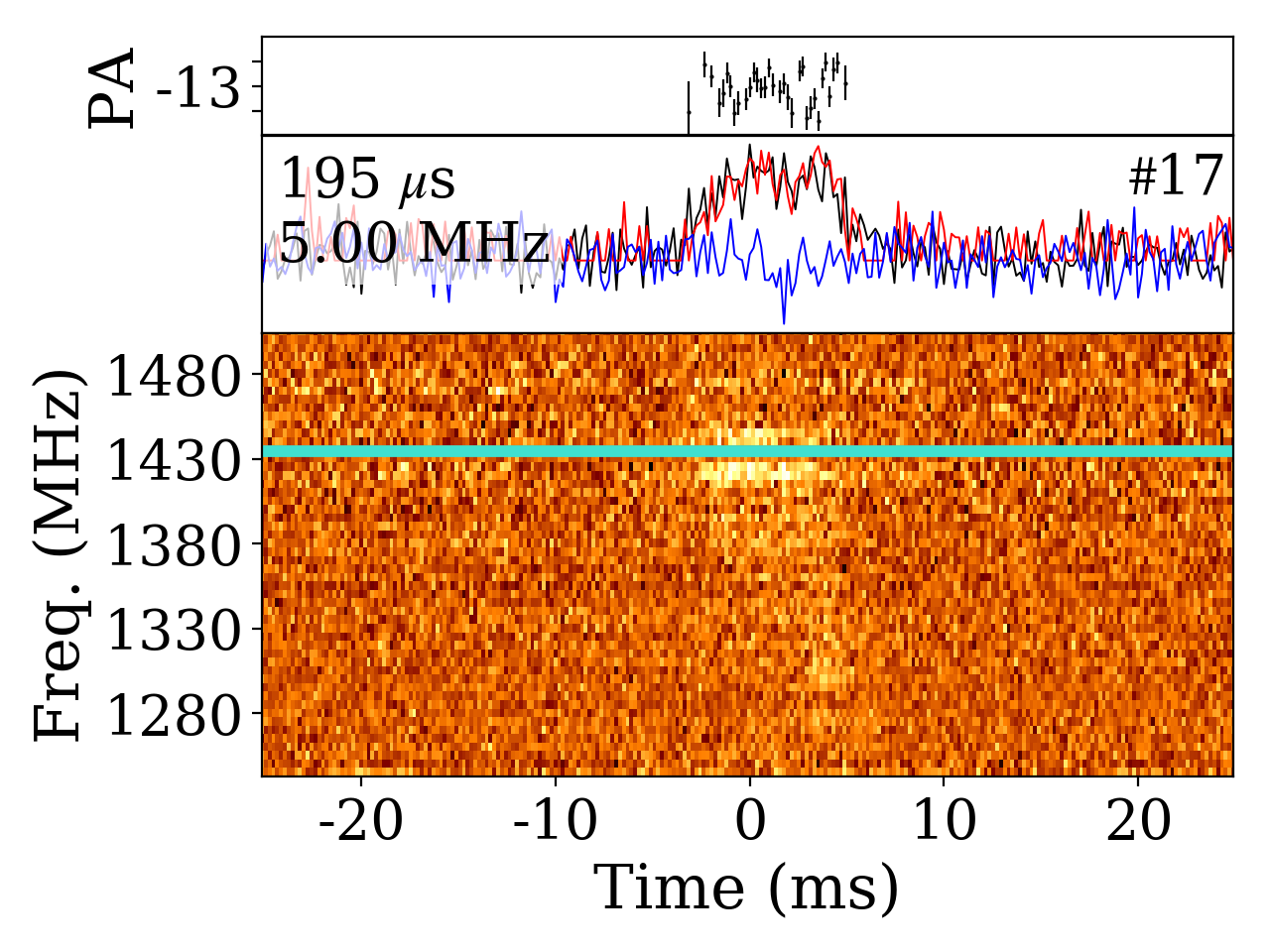}
\includegraphics[width=.24\textwidth]{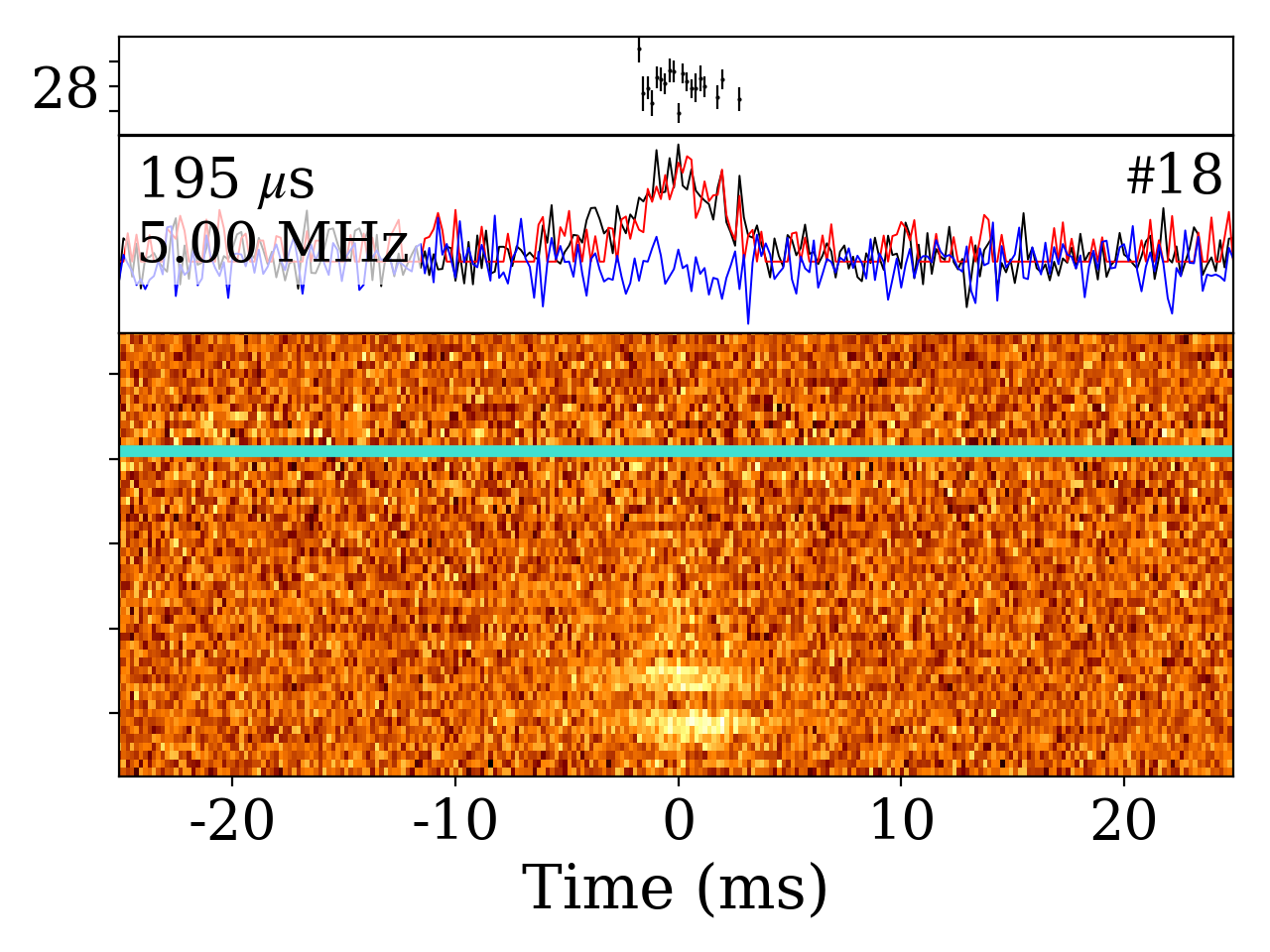}
\includegraphics[width=.24\textwidth]{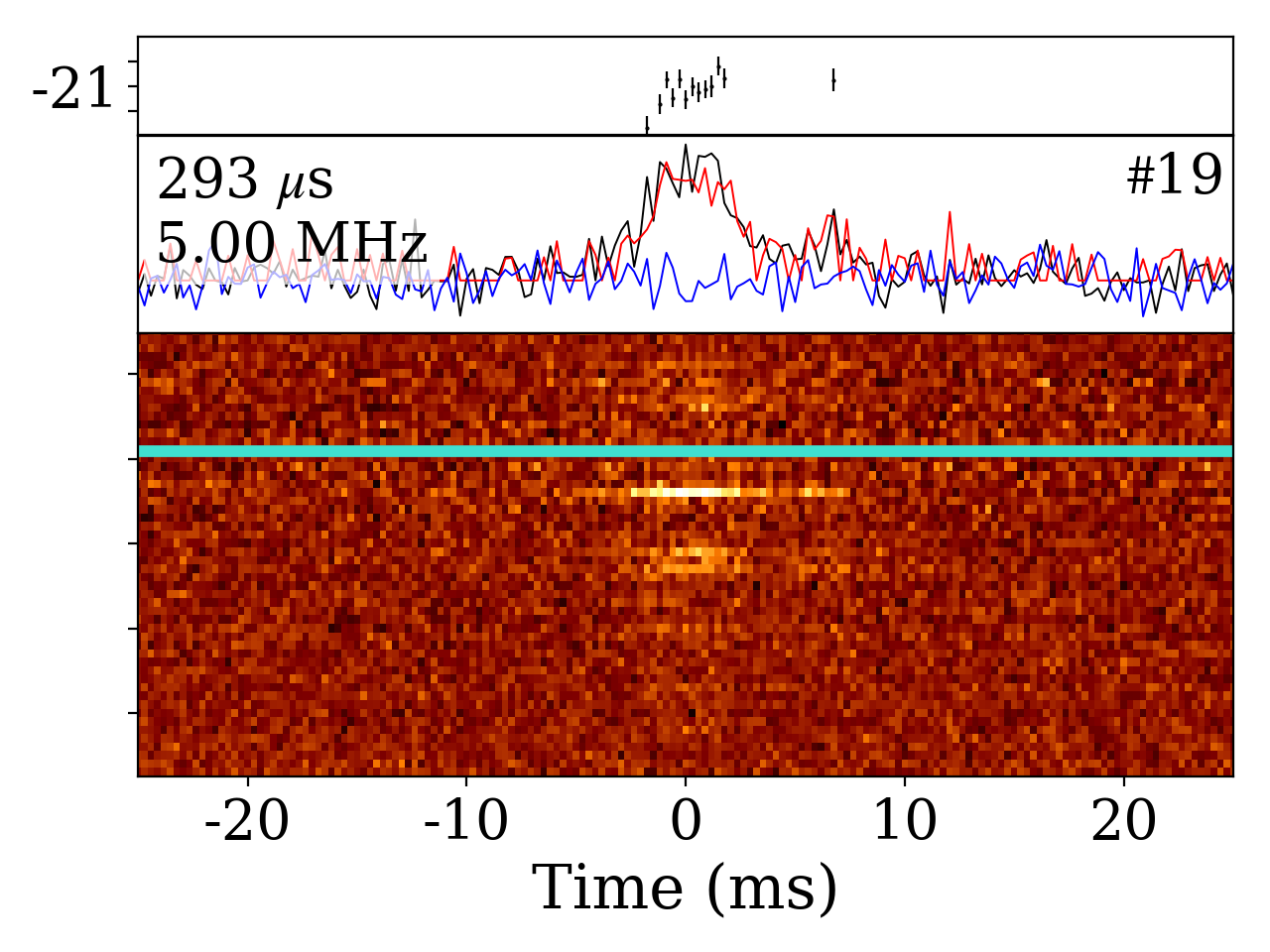}
\includegraphics[width=.24\textwidth]{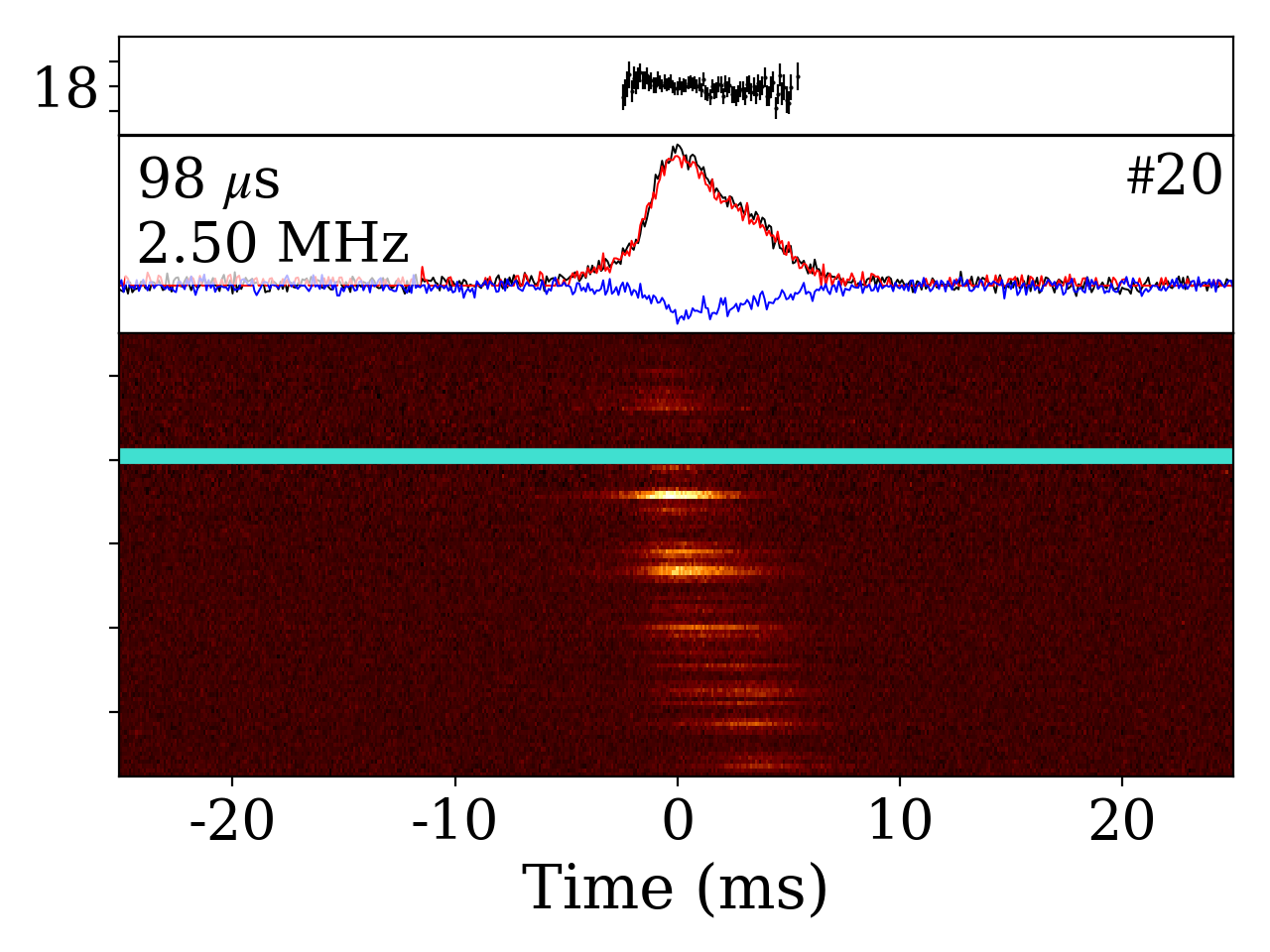}
\caption{
		Dynamic spectra of the bursts. 
		Turquoise horizontal strips are zapped channels. 
		On top of each spectrum is the profile of the burst
		in total intensity (black), unbiased linear polarization (red),
		and circular polarization (blue), and the PA.
		The PA panel range is 40 degrees around the median
		PA value (shifted by $-90^\circ$) of each burst. 
		The time and frequency resolution of each burst is 
		written to the left of its profile, as they are 
		downsampled and subbanded to varying degrees for clarity.
		The burst number (as in Table \ref{tab:bursts}) is to the right of
		each profile.
		}
\label{fig:bursts}
\end{figure*}

We have detected 20 bursts from \frb{} during our observations.
The properties of the bursts are listed in Table \ref{tab:bursts}. 
We fit a Gaussian, and multiple Gaussians in the case of multiple
components, to the profiles of each burst and sub-burst.
The burst widths are determined from the full-width-half-maximum
between the left and rightmost Gaussians.
The flux density and fluence are then determined from the widths and receiver
parameters using the radiometer equation. 
The dynamic spectra, intensity and polarization profiles, and
polarization position angles (PAs) are shown in Fig. \ref{fig:bursts}.
The dynamic spectra in Fig. \ref{fig:bursts} show clear signs
of scintillation that are investigated in \citet{2021arXiv210800052M}.

\begin{table*}
\centering
\caption{
		Burst properties. \textit{From left to right:}
		Burst number, barycentric arrival time in MJD, time in minutes since 
		arrival time of burst 1, width, flux density, fluence, RM, 
		circular polarization fraction, unbiased linear polarization fraction, drift rate, 
		and PA (shifted by $-90^\circ$ to be distributed around zero).
		Uncertainties are 1$\sigma$. 
		The listed PAs are the average PA of each burst weighted by their errors and the uncertainties
		are the standard deviation of the PA values.  
		Burst 11 is considered a single event due to the baseline between the peaks being higher than
		the off-pulse baseline. However, due to the separation of the peaks we estimate a width,
		flux density, and fluence for each one.
		}
\label{tab:bursts}
\begin{tabular}{c|c|c|c|c|c|c|c|c|c|c}
Burst 	& MJD 			& $\Delta T$ & $W$ 	& $S$ 	& $F$ 		& RM & $V/I$ & $L_\mathrm{unb}/I$ & $D$ & PA \\ 
		& (59313$+$) 				& (min) & (ms)	& (Jy)	& (Jy~ms)	& (rad~m$^{-2}$) & $(\%)$ & $(\%)$ & (MHz/ms) & (Deg)\\ \hline	
1	&	0.74325197	& 0 &$6.2\pm0.5$	&$0.17\pm0.03$	& $1.06$		& $-608.5\pm8.6$ & $6\pm7$ & $56\pm5$ & $-41.7\pm1.8$ & $24\pm9$ \\
2	&	0.74897156	& 8.24&$19.3\pm0.4$	&$0.48\pm0.07$	& $9.27$		& $-607.4\pm0.6$ & $-11\pm1$  & $92\pm1$ & $-19.4\pm0.1$ & $5\pm6$ \\
3	&	0.75327832	& 14.44&$15.5\pm0.2$	&$0.23\pm0.03$	& $3.54$		& $-601.9\pm0.4$ & $-1\pm1$ & $94\pm1$ & $-34.9\pm0.1$ & $28\pm4$ \\
4	&	0.76667874	& 33.73&$9.0\pm0.9$	&$0.10\pm0.02$	& $0.92$		& $-599.2\pm4.4$ & $3\pm4$  & $90\pm5$ & $-51.1\pm2.1$ & $24\pm11$ \\
5	&	0.76668105	& 33.74&$8.5\pm0.3$	&$0.50\pm0.07$	& $4.28$		& $-620.4\pm4.2$ & $-20\pm2$ & $99\pm2$  & $-52.0\pm0.4$ & $-6\pm5$ \\
6	&	0.77488410	& 45.55&$22.2\pm1.6$	&$1.27\pm0.19$	& $28.12$		& $-608.3\pm0.3$ & $6\pm1^{\mathrm{a}}$ & $98\pm1$ & $-16.5\pm0.1$ & $35\pm2$ \\
7	&	0.78795702	& 64.38&$5.6\pm0.6$	&$0.40\pm0.06$	& $2.22$		& $-604.6\pm2.5$ & $3\pm2$ & $97\pm2$ & $-88.6\pm0.6$ & $41\pm6$ \\
8	&	0.80949925	& 95.40&$6.4\pm0.5$	&$0.16\pm0.02$	& $1.02$		& $-612.0\pm5.8$ & $0\pm4$ & $90\pm4$ & $-51.4\pm0.8$ & $-6\pm7$ \\
9	&	0.81914765	& 109.29&$10.3\pm0.3$	&$2.31\pm0.35$	& $23.78$		& $-584.1\pm0.3$ & $-5\pm1$ & $99\pm1$ & $-29.0\pm0.1$ & $10\pm2$ \\
10	&	0.82721002	& 120.90&$7.1\pm1.1$	&$0.09\pm0.01$	& $0.66$		& $-604.7\pm11.7$ & $-3\pm9$ & $78\pm8$ & $-45.8\pm2.8$ & $-27\pm11$ \\
11a	&	\multirow{2}{*}{0.82721037}	& \multirow{2}{*}{120.90}&$7.8\pm0.8$	&$0.06\pm0.01$	& $0.47$		& \multirow{2}{*}{$-601.2\pm11.7$} & \multirow{2}{*}{$7\pm4$} & \multirow{2}{*}{$84\pm4$} & \multirow{2}{*}{$-108.3\pm11.1$} & \multirow{2}{*}{$13\pm11$} \\
11b	&		& &$7.1\pm1.0$	&$0.07\pm0.02$	& $0.47$		&  &  &  & & \\
12	&	0.83062730	& 125.82&$11.1\pm0.3$	&$0.10\pm0.01$	& $1.09$		& $-618.0\pm1.4$ & $-3\pm1$ & $90\pm1$ & $-28.5\pm0.1$ & $9\pm6$ \\
13	&	0.83240637	& 128.38&$12.1\pm0.7$	&$0.06\pm0.01$	& $0.76$		& $-599.4\pm2.7$ & $10\pm4$ & $100\pm3$ & $-32.3\pm0.5$ & $-1\pm8$ \\
14	&	0.84107824	& 140.87&$9.7\pm0.8$	&$0.12\pm0.02$	& $1.19$		& $-600.5\pm6.3$ & $3\pm5$ & $91\pm4$ & $-28.9\pm0.3$ & $-3\pm10$ \\
15	&	0.84379515	& 144.78&$4.8\pm0.1$	&$0.57\pm0.09$	& $2.75$		& $-614.7\pm1.3$ & $-6\pm2$ & $100\pm2$ & $-65.2\pm0.2$ & $6\pm4$ \\
16	&	0.84713858	& 149.60&$10.4\pm0.3$	&$0.09\pm0.01$	& $0.94$		& $-616.9\pm1.3$ & $4\pm1$ & $96\pm2$ & $-23.7\pm0.1$ & $6\pm7$ \\
17	&	0.85342708	& 158.65&$7.5\pm1.5$	&$0.12\pm0.02$	& $0.91$		& $-601.8\pm4.4$ & $-3\pm4$ & $96\pm4$ & $-59.9\pm1.2$ & $-13\pm7$ \\
18	&	0.85674812	& 163.43&$7.4\pm0.5$	&$0.14\pm0.02$	& $1.02$		& $-579.5\pm6.7$ & $-17\pm5$ & $84\pm4$ & $-46.0\pm1.6$ & $29\pm6$ \\
19	&	0.88723065	& 207.33&$4.7\pm0.4$	&$0.14\pm0.02$	& $0.63$		& $-592.5\pm5.0$ & $0\pm4$ & $92\pm4$ & $-94.0\pm5.7$ & $-22\pm6$ \\
20	&	0.89050534	& 212.04&$5.9\pm0.1$	&$1.51\pm0.23$	& $8.94$		& $-624.0\pm0.6$ & $-21\pm1$ & $96\pm1$ & $-45.9\pm0.1$ & $18\pm3$ \\
\multicolumn{10}{l}{\footnotesize{$^\text{a}$Absolute Stokes $V$ values used}}\\
\multicolumn{10}{l}{\footnotesize{Removing the baseline can result in $I>=\sqrt{Q^2+U^2+V^2}$ not being fulfilled}}
\end{tabular}
\end{table*}

\subsection{Periodicity and Burst Rate}

Using a Lomb-Scargle periodogram we search for periodicities
in our sample.
We detect no periodicity in burst arrival times from the 
minimum time difference between bursts in our sample up to the 
full 4 hours of our observation. 
This is not surprising given the periodicities of \Ri{} and \Riii{}
of 10s/100s days, and no short time scale periodicity ($P<1$s) has been determined for any repeating FRB 
\citep[e.g.][]{2020arXiv201208348P,2021MNRAS.500..448C}. 

The average burst rate during our observing window is $5.0^{+1.4}_{-1.1}$~bursts/hr
above a fluence of $0.14\times(W_\mathrm{ms})^{1/2}$~Jy~ms
for an S/N threshold of 7. 
We estimate a Weibull distribution clustering parameter of
$k=0.7\pm0.1$ using the full sample. If bursts 10 and 11,
separated by 30~ms, are consolidated into a single event, we obtain $k=0.9\pm0.2$.
A clustering parameter of $k=1$ reduces a Weibull distribution
to a Poissonian one. 
Therefore, the arrival times of \frb{} during this observation 
are consistent with a Poisson distribution when bursts separated 
by $\lesssim 100$ms are excluded, 
as has also been seen in \Ri{} \citep{2021MNRAS.500..448C}. 

\subsection{DM \& RM}

We estimate a singular DM that we apply to all our bursts.
Using the results of \texttt{DM\_phase} of the multi-component
bursts 2, 3, 6, 11, and 12, we obtain an average DM and error of 
\DM{$411.6\pm0.6$}. 

According to the NE2001 Galactic electron density model \citep{2002astro.ph..7156C}
the Galactic DM contribution along the line of sight towards
\frb{} is \DM{$123\pm25$}
\citep[20\% accuracy, e.g.][]{2009ApJ...701.1243D}.
For YMW16 the Galactic DM contribution is \DM{197}
\citep{2019ascl.soft08022Y}.
We assume a DM of \DM{50--80} from the Galactic halo \citep{2019MNRAS.485..648P}.
The estimated redshift of the host galaxy of \frb{}
is $0.098\pm0.002$ \citep{2021ATel14516....1K}, 
and the intergalactic medium (IGM)-redshift relation
from \citet{2020Natur.581..391M} yields an IGM DM contribution of \DM{$82^{+17}_{-30}$} ($1\sigma$).
The resulting host galaxy DM is thus \DM{$140^{+33}_{-42}$}.

\begin{figure}
\centering
\includegraphics[width=\columnwidth]{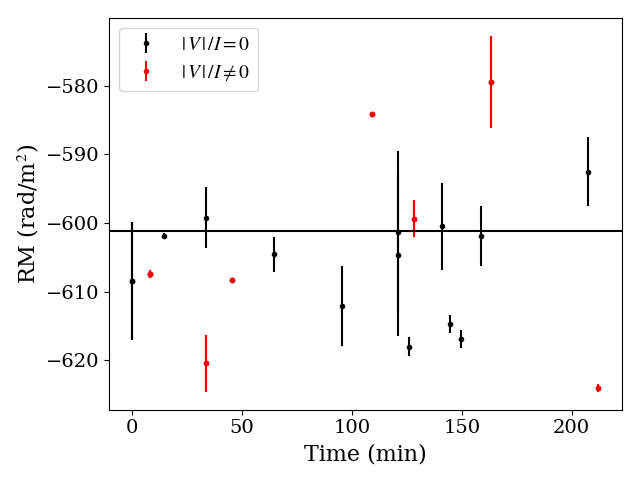}
\includegraphics[width=\columnwidth]{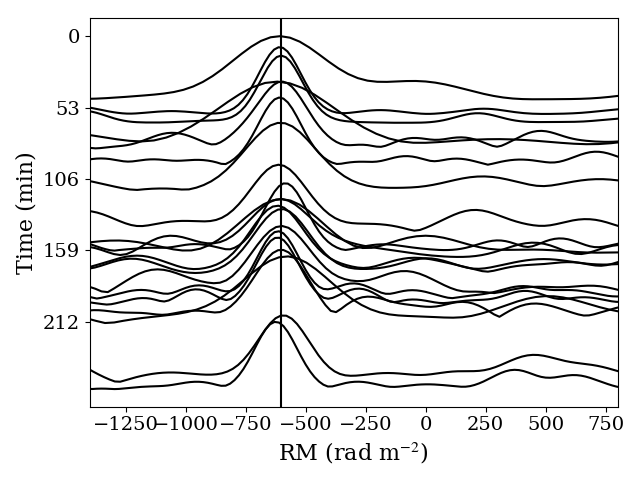}
\caption{
		\textit{Top}:
		RM over time of detected bursts. The horizontal line
		is the weighted mean RM of all the bursts, \RM{-601}.
		Black points are bursts with circular polarization
		fraction consistent with noise, red points are bursts
		with non-zero circular polarization fractions.
		\textit{Bottom}:
		FDFs of the bursts, scaled to the same height.
		Peaks of the FDFs are at a time in minutes relative to the 
		first burst as in Table \ref{tab:bursts}.
		The vertical line shows the weighted mean RM value of the bursts.
		}
\label{fig:fdf}
\end{figure}

The RMs we obtain for our bursts are listed in Table \ref{tab:bursts}
and plotted over time in minutes with respect to the first burst
in the upper panel of Fig. \ref{fig:fdf}. We also plot the weighted
mean RM value of \RM{-601}. 
The lower panel of Fig. \ref{fig:fdf} shows the FDF of our bursts.
The standard deviation of the burst RMs is \RM{11.1}, which is a
fractional variation of 1.8\% from the mean.

From the coordinates of \frb{} we can calculate the Galactic RM
contribution\footnote{\url{github.com/FRBs/FRB}} based on the
Galactic RM foreground map from \citet{2015A&A...575A.118O},
and obtain a value of \RM{$-57\pm33$}. 
We assume that the IGM has a negligible contribution to the RM 
\citep[e.g.][]{2015A&A...575A.118O}.
The majority of the measured RM must therefore originate from
the host galaxy and local environment of \frb{}
which amounts to an observed RM of \RM{$-548\pm35$}, or
\RM{$-661\pm42$} in the redshifted source frame.

\subsection{Polarimetry}

All our bursts exhibit a high linear polarization fraction,
consistent with linear polarization of other repeating FRBs
\citep[e.g.][]{2018Natur.553..182M,2020ApJ...896L..41C,2021ApJ...908L..10H,2021NatAs.tmp...53N}.
For bursts 2, 5, 6, 9, 13, 18, and 20 we find evidence of 
circular polarization, ranging from absolute Stokes $V/I$
fractions of 6--21\%.
For the other bursts the circular polarization is consistent with noise,
i.e.\ the Stokes $V$ on/off-pulse standard deviation does not differ.
Linear and circular polarization fractions are listed in Table \ref{tab:bursts}.
The measured linear polarization, $L=\sqrt{Q^2+U^2}$, is overestimated in 
the presence of noise. We therefore use the unbiased 
linear polarization \citep{2001ApJ...553..341E}
\begin{equation}
\label{eq:Lunb}
L_\mathrm{unb} = \begin{cases}
\sigma_I \sqrt{\left( \frac{L}{\sigma_I} \right)^2 - 1} & \mathrm{if~} \frac{L}{\sigma_I} \geq 1.57 \\
0 & \mathrm{otherwise,}
\end{cases}
\end{equation}
where $\sigma_I$ is the Stokes $I$ off-pulse standard deviation.
Henceforth all mention of linear polarization refers to $L_\mathrm{unb}$.

We split our band into 4 subbands to see if the circular polarization is localized to specific parts of the band. Most of the bursts do not cover the entire band, so both Stokes $I$ and $V$ are only seen in some of the subbands. However, burst 20 covers most of our band and has a relatively high circular polarization fraction that is consistent across all subbands. 

\begin{figure}
\centering
\includegraphics[width=\columnwidth]{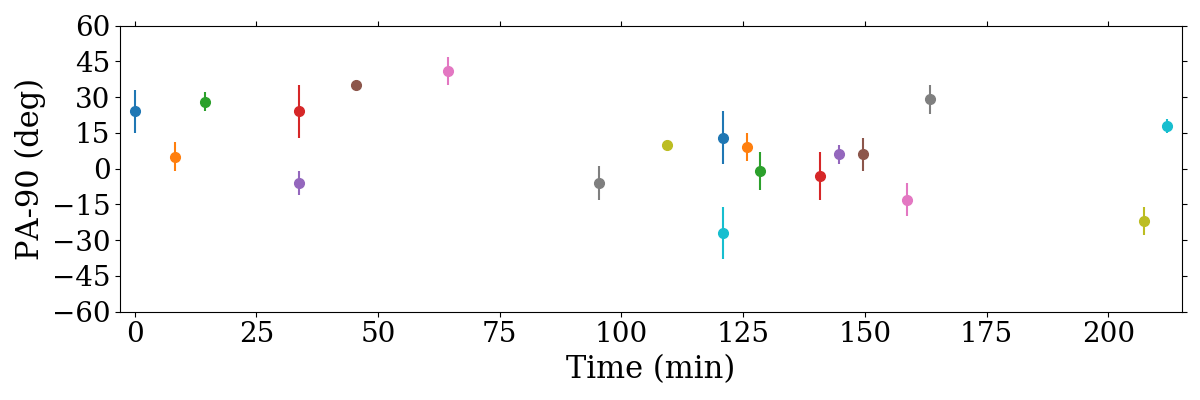}
\includegraphics[width=\columnwidth]{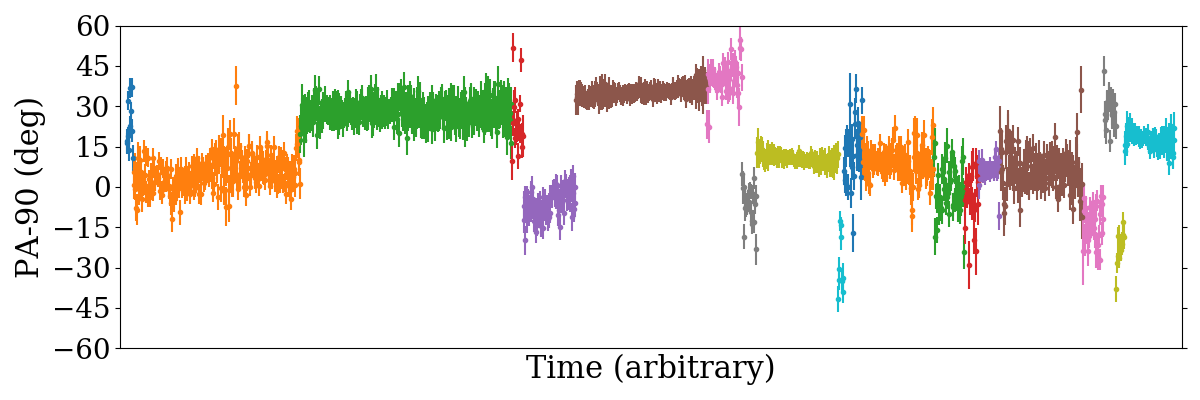}
\caption{
		\textit{Top:}
		Weighted average PA of each burst over time (shifted by $-90^\circ$)
		as listed in Table \ref{tab:bursts}.
		\textit{Bottom:}
		PAs across each burst envelope as in Fig. \ref{fig:bursts} shown
		sequentially.
		}
\label{fig:PAtime}
\end{figure}

The PAs are flat across the bursts, but show slight variations 
between bursts. The mean of the medians of the burst PAs is
$8^\circ$ with a standard deviation of $18^\circ$.  
Fig. \ref{fig:PAtime} shows the PA over time of our bursts relative
to their arrival time in minutes (top panel) and sequentially (bottom panel).
If all the bursts are derotated to the average RM of \RM{-605} we find the burst-to-burst PA variation slightly larger than when the bursts are derotated to their specific RM value.
Our PAs are calculated with respect to a reference angle at
infinite frequency ($\mathrm{PA}=\mathrm{RM}\lambda^2 + \mathrm{PA}_\infty$).

\subsection{Spectro-temporal Properties}

\begin{figure}
\centering
\includegraphics[width=\columnwidth]{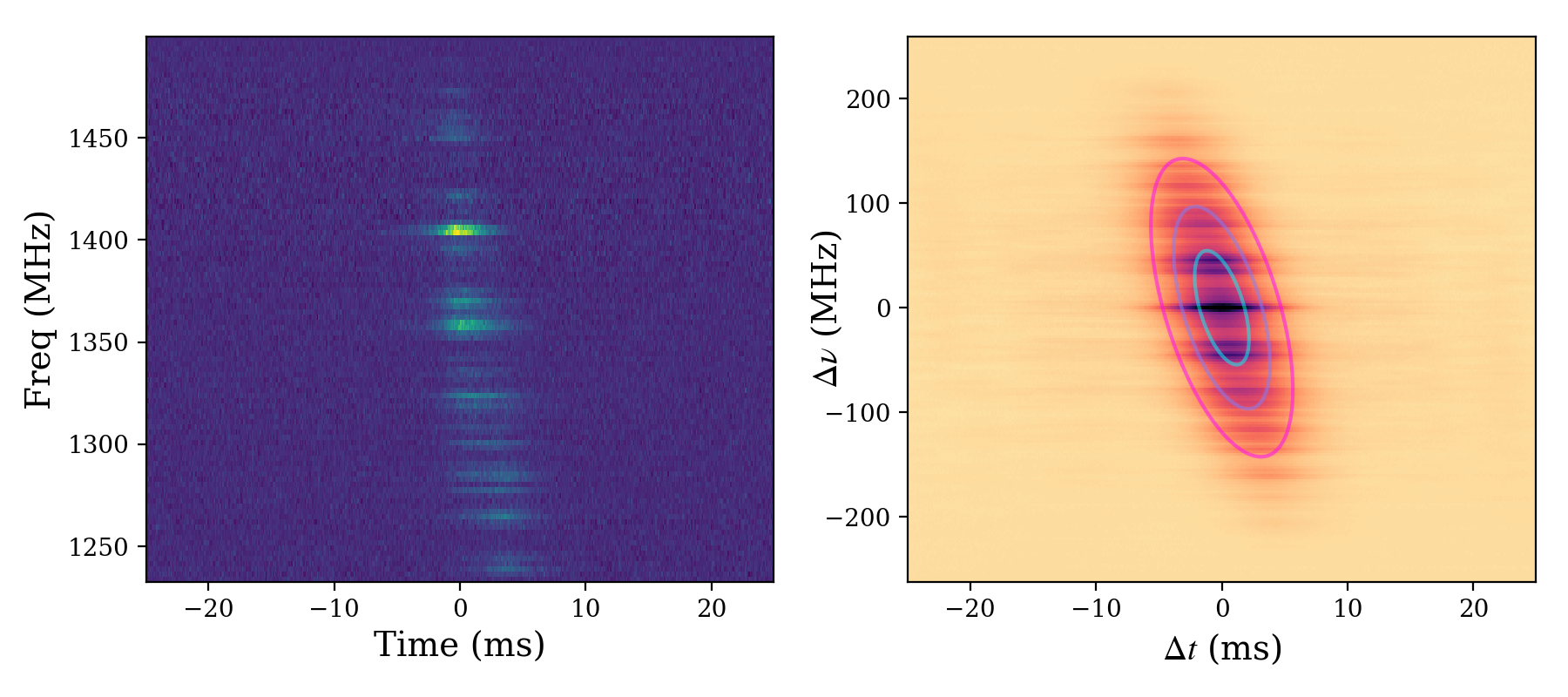}
\caption{
		Drift analysis of burst 20.
		\textit{Left:} Dynamic spectrum of the burst, 
		dedispersed to \DM{411.6},
		with the same time and frequency binning as in Figure \ref{fig:bursts}.
		\textit{Right:} 2D ACF of the burst.  Contours show the 1, 2, and 3$\sigma$ widths of the best-fit 2D Gaussian, used to derive the drift rate. The fine frequency structure arises from the scintillation, and is unrelated to the downward drift.
		}
\label{fig:drift}
\end{figure}

A downward drift in frequency can clearly be seen for many of 
our bursts in Fig. \ref{fig:bursts}. We quantify the drifts using
a 2D autocorrelation function.  We perform a 2D autocorrelation of the 
intensity of each burst, using the same time and frequency ranges and 
resolutions as Fig. \ref{fig:bursts}.  We fit the result with a 2D rotated 
Gaussian, from which we obtain the burst drift rate; the results are listed 
in Table \ref{tab:bursts}, and an example output of the drift analysis for 
burst 20 is shown in Fig. \ref{fig:drift}.

Our burst widths, with a mean value of 9.5~ms, are unusually large when
compared to other repeating FRBs observed at 1.4 GHz
\citep[e.g.][]{2019ApJ...876L..23H,2019ApJ...885L..24C,2020ApJ...891L...6F}.
These widths are likely intrinsic to the burst, as scattering inferred from the scintillation bandwidth of these
bursts is $\tau_{\rm 1\,GHz} = 3.1 \pm 0.06\; \mu$s 
\citep{2021arXiv210800052M} 
and we do not observe a consistent 
asymmetry in the trailing edges of the burst profiles that would be consistent with scattering.

Burst 11 is considered a single burst here due to the fact that there
are only a few milliseconds between its peaks and that the baseline
between the peaks is higher than the off-pulse baseline. We still 
however split the burst into sub-bursts \textit{a} and \textit{b}
with separate widths.

\section{DISCUSSION}\label{sect:disc}

\frb{} is the first repeating FRB to show clear signs of circular polarization,
with circular polarization fractions up to 20\% in roughly one-third of the sample. We note that the bursts with non-zero circular polarization occur throughout the observation and never occur consecutively. This suggests the mechanism is time variable on short time scales. There is no correlation between non-zero circular polarization and PAs, which one may expect if the bursts originate in a neutron star magnetosphere. In Figure~\ref{fig:fdf} one can see that four of the  bursts with non-zero $V/I$ have the four most extreme RM values (bursts 5, 9, 18, 20). 

The high linear polarization fractions, flat PAs across burst envelopes, and drift rates of \frb{} are consistent with
what has been observed in most other repeating FRBs including 
\Ri{}, \Riii{}, and \Rbode{} 
\citep{2018Natur.553..182M,2019ApJ...876L..23H,2020Natur.582..351C,2021ApJ...908L..10H,2021NatAs.tmp...53N,2021arXiv210511446N}. The exception is \Rfast{}, which exhibits variable, linear polarization $<100\%$ and a drifting PA across the burst envelopes but no circular polarisation \citep{2020Natur.586..693L}.
\frb{} is, therefore, similar in many way to other repeating FRBs, but its non-zero circular polarization observable in some, but not all, makes the observed properties of repeaters even more diverse. 

Comparisons are often made between FRBs and pulsar giant pulses (GPs), particularly those from the Crab pulsar. \citet{2019ApJ...876L..23H} noted the similarity in the banded structure observed in the spectra of Crab giant pulses detected at higher radio frequencies ($\gtrsim 5$~GHz) from the interpulse (so called HFIP GPs) and the band-limited emission in repeating FRBs. \citet{jessner+2010} measured the polarization properties of a large sample of giant pulses at 8.5~GHz and 15.1~GHz from both the main pulse (MP) and interpulse (IP). A large fraction of the GPs from both the MP and IP have 100\% linear polarization, but pulses with circular polarization fractions between a few -- of order 10\% were not uncommon. What is markedly different between the MP and IP GPs is the variation in PA. While PAs for the MPs were varied between 0 and 180 degrees, the PAs for IPs were more restricted in range with $\sim80\%$ of the bursts having PAs that varied within 30 degrees \citep{jessner+2010}. Figure~\ref{fig:PAtime} shows that the variation in the PAs from \frb{} is similar, with most PAs falling within a 30 to 40 degree range. Combined with the banded structure seen in the dynamic spectra, bursts from \frb{} show striking similarities to the HFIP GPs from the Crab pulsar. 

Polarization observations of the magnetar XTE J1810-197 
\citep{2007MNRAS.377..107K}
also show noticeable similarities to \frb{}.
The profiles of XTE J1810-197, both single pulse and
averaged over multiple epochs, have high degrees of
linear polarization and low but significant degree
of circular polarization. The PAs are also slightly slanted
across the profile, which is more akin to what we see
from \frb{} and other repeating FRBs rather than the 
S-like swing seen from pulsars.

In magneto-active plasma, linearly polarized radiation can be converted into circular in a process called ``Faraday conversion" or ``generalized Faraday rotation". \cite{2019MNRAS.485L..78V,2019ApJ...876...74G} have discussed this effect on \Ri{} for relativistic and non-relativistic case respectively. However, it is challleging for this effect to account for the change of circular polarization of \frb{}. First, the change of RM is only $\sim10$~rad/m$^2$ for the bursts with circular polarization. Given a host DM of $\sim 100$~pc/cm$^3$, it corresponds to magnetic field strength of only $\mu$G with the cyclotron frequency $f_B=2.8\mathrm{MHz/G}\,B\sim$Hz. In this case, $f_B$ is 9 orders of magnitude lower than the observation frequency at GHz, while Faraday conversion usually happens when the cyclotron frequency is approaching the observation frequency. Even if only a small fraction of DM is contributing to the RM, and the B field is several orders larger, it is difficult to bridge the 9 order difference, otherwise the local electron density would be too low and the electric field would again not experience the Faraday conversion. Second, bursts separated by minutes are seen to have different circular fractions, which gives an estimate of the spatial scale of changes assuming the typical velocity of few hundred km/s for neutron stars. This rules out the scenario of Faraday conversion with weak magnetic field,  where the EM wave adiabatically tracking a slow field reversal in the LOS \citep{2019ApJ...876...74G}. On the other hand, if there is a group of highly-relativistic electron introducing the Faraday conversion, the circular and linear rotation can be modeled with RRM$\lambda^3$\citep{1998PASA...15..211K}. The limited bandwidth has prevented us from distinguishing $\lambda^2$ rotation from RM and $\lambda^3$ rotation from RRM and hence the observed RM variation could be due to RRM, which will introduce various levels of circular polarization depending on the local magnetic field, electron density and the angle between field line and polarization angle. However, assume the field to be fully linearly polarized at infinite frequency, given the observed low circular fraction, any Faraday conversion has to be incomplete. Therefore the fraction in the lower band would be higher than the upper band. With $\lambda^3$ dependence, the polarization fraction at the bottom of the band should be $\sim1.5$ times the polarization fraction at the top of the band, which is not seen in the data. Therefore, we conclude that the observed circular polarization may be intrinsic to the source.

In terms of RM magnitude among repeating FRBs, \frb{} is
similar to FRB 20190303A and \Rfast{}\ \citep{2020ApJ...891L...6F,2020Natur.586..693L},
while others seem to be more or less consistent with the
Galactic RM contribution 
\citep[\Riii{}, FRB 20190604A, FRB 20200120E][]{2019ApJ...885L..24C,2020ApJ...891L...6F,2021arXiv210511446N}.
Finally, there is \Ri{} with its enormous RM of \RM{$\sim10^5$}
\citep{2018Natur.553..182M,2021ApJ...908L..10H}.

Similar temporal RM variability is seen in \Rfast{} \citep{2020Natur.586..693L}
where its RM varies between \RM{520--560} over a 1-day timescale
and is thought unlikely to be a constant.
Another example is the RM variation of \Ri{},
showing RM differences of \RM{$\sim10^2$} 
over 1-hour timescales \citep{2018Natur.553..182M}.
However, in that case a global RM average is calculated for
each observing epoch.
In this context, \Riii{} has no short timescale measurements
but has been observed to be consistent with its original RM
value of \RM{-115} \citep{2021NatAs.tmp...53N} and seen to decrease by only a
few rad~m$^{-2}$ over the course of a year \citep{2021ApJ...911L...3P}.
Further polarization measurements of \frb{} will determine the stability of its RM.

Assuming that all the host DM is contained within the Faraday rotating
medium, we can estimate a source frame average magnetic field strength
along the line of sight as
\begin{equation}
\langle B_\parallel \rangle = 1.23\, \mathrm{RM}_\mathrm{src}/\mathrm{DM}_\mathrm{host} \mathrm{~}\mu\mathrm{G},
\end{equation}
and obtain an estimate of 4--6~$\mu$G. If the RM originates in a small physical region close to the source, such as the remnants of a supernova, the amount of DM in this Faraday rotating medium could be much smaller, and the above estimate is a lower limit.

RM variations on a time scale of minutes must arise from spatial variations in the electron density and/or line-of-sight magnetic field strength close to the source. The observed fractional RM variations are roughly 2\%. Assuming a constant DM, the magnetic field strength variations must be on the order of 0.1~$\mu$G.
If instead the observed RM variation
is due to changes in the host DM contribution, the host DM must vary 
by \DM{2--4}, which translates to a fractional variation of the 
total observed DM of $0.7\%$. 
This change is inconsistent with our lack of change in DM between bursts, suggesting that either the RM variations are due to small-scale magnetic field variations or only a fraction of the estimated host DM contributes to the RM. 

We also note that the RM measurement uncertainties of \frb{} could also be underestimated and the RM value is, in fact, constant with time. 
The FDFs in the lower panel of Fig. \ref{fig:fdf} show that the
mean RM is well contained within their widths. 

We can consider the model of a magnetar within a supernova remnant 
from \cite{2018ApJ...868L...4M}, 
where bursts are generated by a synchrotron maser
due to shocks from the magnetized outflow of the magnetar into its
circumsource material.
The RM generating material is contained within the expanding magnetar
nebula, and for our mean RM magnitude of \RM{601} we obtain source
ages between 60--200 years.
At this age, the coincident, compact PRS should be at the $\mu$Jy level,
which is consistent with the non-detection of such a source associated
with \frb{} \citep{2021ATel14603....1M,2021arXiv210609710R}.
Within this framework the source ages of \Ri{} and \Riii{}
have been estimated to be $\sim10$ and $\sim300$ years, respectively
\citep{2020Natur.577..190M,2021ApJ...908L..10H}.

Prior to being found to be periodic in its activity, bursts from
\Ri{} appeared to follow a Weibull distribution, i.e. clustered in
time with a Weibull clustering factor of $k=0.34$ 
\citep{2018MNRAS.475.5109O}. 
Within a typical observation time of a few hours, bursts from \Ri{} follow
a Poissonian distribution \citep{2021MNRAS.500..448C}.
If \frb{} is periodic in its activity, one can expect burst arrival times to follow a Weibull distribution over long time scales until a periodicity is found.

As mentioned in Section~\ref{sect:results}, the mean and narrowest burst durations of our sample are 9.5~ms and 4.7~ms, respectively. By comparison, the peak of the width distribution of $\sim$1600 bursts from \Ri{} is $\sim$4~ms \citep{2021arXiv210708205L}. The widths of four bursts detected from \Rbode{} range from 50 to 150~$\mu$s \citep{2021arXiv210511446N}. Therefore, the characteristic burst duration varies over two orders of magnitude among the known repeating FRBs. If bursts from repeating FRBs originate in the magnetosphere of neutron stars, the characteristic duration may correlate with rotational period, as seen in micro-structure of pulsar single pulses \citep[e.g.][]{2002MNRAS.334..523K}. 

\begin{figure}
\centering
\includegraphics[width=\columnwidth]{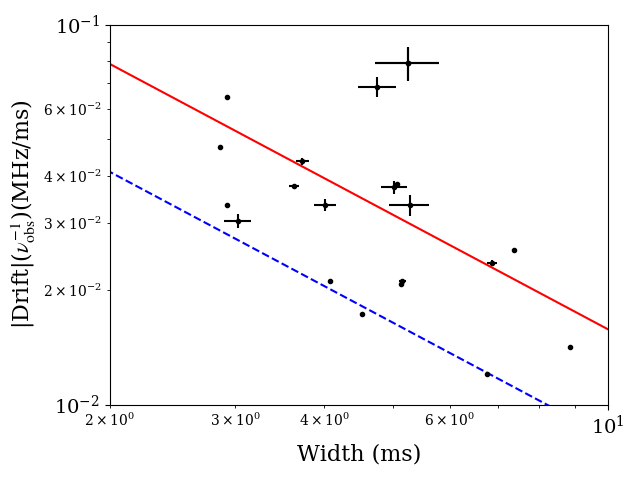}
\caption{
		Best-fit 2D Gaussian widths versus absolute burst drifts
		scaled by the central observing frequency.
		Red line shows the best fit through the data and
		the blue dashed line shows the results from 
		\citet{2020arXiv201014041C}.
		}
\label{fig:WvsD}
\end{figure}

\citet{2020arXiv201014041C} investigate the relation between burst widths and drift rates
within the framework of Dicke's superradiance, and find a linear relation
between the two.
We plot the burst widths versus drift for our burst sample in Fig. \ref{fig:WvsD}.
To be consistent with the methods in \citet{2020arXiv201014041C} we
use the $1\sigma$ widths of the fitted rotated 2D Gaussians (right panel in Fig.
\ref{fig:drift}) projected onto the time axis.
A best fit line through our data points yields a similar slope to
the results of \citet{2020arXiv201014041C}, but slightly offset
towards higher widths.

\section{SUMMARY}\label{sect:summ}

We have detected 20 bursts from \frb{} from
a single 4-hour observation on April 9 2021
with the Effelsberg 100-m radio telescope at 1.4~GHz.
As commonly seen in repeating FRBs, \frb{}
exhibits a downward drift in frequency over time
in many of its bursts.
However, the burst widths are wider than normally
seen at this frequency, with a mean value of 9.5~ms.
The bursts in our sample are highly linearly polarized
($\gtrsim 75\%$),
with a subset of them also being circularly polarized
($\lesssim 20\%$) which is a first for a repeating FRB.
We obtain an average DM of \DM{$411.6\pm0.6$}.
The RM of \frb{} varies slightly between bursts
around their weighted mean of \RM{-601}.
Interestingly, the bursts with circular polarization
have the most extreme RM values in our sample.
We hope that future polarization measurements of
\frb{} will help determine its RM stability.

\section{ACKNOWLEDGEMENTS}
Based on observations with the 100-m telescope of the MPIfR (Max-Planck-Institut f\"{u}r Radioastronomie) at Effelsberg.
Part of this research has made use of the EPN Database of Pulsar Profiles maintained by the University of Manchester, available at: jodrellbank.manchester.ac.uk/research/pulsar/Resources/epn/.
GHH thanks Dr. A.~D.~Seymour for insights into DM-Phase, Dr. N.~K.~Porayko for 
help with RM search methods and archive handling, and Dr. M.~Cruces for providing
helpful code snippets.
LGS is a Lise Meitner Max Planck independent research group leader and acknowledges funding from the Max Planck Society. 
We thank the anonymous referee for their helpful comments.

\section{DATA AVAILABILITY}
The data underlying this article will be shared on reasonable request to the corresponding authors.

\bibliographystyle{mnras}
\interlinepenalty=10000
\bibliography{bibby}

\end{document}